\documentclass[aps,pra,twocolumn,floatfix,superscriptaddress]{revtex4}

\usepackage{amssymb,amsmath,amstext}                
\usepackage{graphicx}                                               
\usepackage{epstopdf}                                               
\usepackage{color}                                                     
\usepackage{bm}                                                        
\usepackage{appendix}                                              
\usepackage[polish]{babel}
\usepackage{bbold}
\usepackage{bbm}    
\usepackage{ulem}
\usepackage{latexsym}
\usepackage[colorlinks=true,citecolor=blue,linkcolor=magenta]{hyperref}
\usepackage{cancel}

\usepackage[dvipsnames]{xcolor}

\def\be{\begin{equation}} 
\def\ee{\end{equation}}
\def\bea{\begin{eqnarray}}
\def\eea{\end{eqnarray}}
\def\ra{\rangle}
\def\la{\langle} 
\def\bi{\begin{itemize}}
\def\ei{\end{itemize}}

\definecolor{dgreen} {RGB}{78,138,21} 
\definecolor{aa} {RGB}{40,153,28}

\begin{document} 

\title{Dynamical quantum phase transitions in systems with broken continuous time and space translation symmetries}

\author{Arkadiusz Kosior} 
\affiliation{Instytut Fizyki imienia Mariana Smoluchowskiego, 
Uniwersytet Jagiello\'nski, ulica Profesora Stanis\l{}awa \L{}ojasiewicza 11, PL-30-348 Krak\'ow, Poland}
\author{Andrzej Syrwid} 
\affiliation{Instytut Fizyki imienia Mariana Smoluchowskiego, 
Uniwersytet Jagiello\'nski, ulica Profesora Stanis\l{}awa \L{}ojasiewicza 11, PL-30-348 Krak\'ow, Poland}
\author{Krzysztof Sacha} 
\affiliation{Instytut Fizyki imienia Mariana Smoluchowskiego, 
Uniwersytet Jagiello\'nski, ulica Profesora Stanis\l{}awa \L{}ojasiewicza 11, PL-30-348 Krak\'ow, Poland}
\affiliation{Mark Kac Complex Systems Research Center, Uniwersytet Jagiello\'nski, ulica Profesora Stanis\l{}awa \L{}ojasiewicza 11, PL-30-348 Krak\'ow, Poland
}

\date{\today}

\begin{abstract}
Spontaneous breaking of continuous time translation symmetry into a discrete one is related to time crystal formation. While the phenomenon is not possible in the ground state of a time-independent many-body system, it can occur in an excited eigenstate. Here, we concentrate on bosons on a ring with attractive contact interactions and analyze a quantum quench from the time crystal regime to the non-interacting regime. We show that dynamical quantum phase transitions can be observed where the return probability of the system to the initial state before the quench reveals a non-analytical behavior in time. The problem we consider constitutes an example of the dynamical quantum phase transitions in a system where both time and space continuous translation symmetries are broken. 
\end{abstract}
\date{\today} 

\maketitle
 
\section{Introduction}

Spontaneous symmetry breaking is related to a phenomenon when equations describing a system possess a certain symmetry, but a system chooses spontaneously a solution which breaks this symmetry. It can be associated with the fragility of exact symmetric eigenstates against infinitesimally weak perturbations.  Spontaneous symmetry breaking is responsible for a wide class of phenomena, e.g., non-zero  magnetization of a ferromagnetic material or formation of space or time crystals \cite{Sacha2017rev}. Usually, spontaneous symmetry breaking is accompanied by a phase transition. That is,  there is a critical value of a control parameter of systems which separates symmetric and symmetry broken phases. 
  
While the equilibrium phase transitions are quite well understood,  the same cannot be said about non-equilibrium dynamics of quantum many-body systems  \cite{Sachdev2011,Dziarmaga2010}. 
Most notably, the pioneering works of Kibble and Zurek \cite{Kibble1980,Zurek1996} led to the discovery of dynamical creation of topological defects in systems driven through a  quantum critical point at a finite rate \cite{Damski2005,Zurek2005,Dziarmaga2005,Polkovnikov2005,Uhlmann2007,Uhlmann2010,Swislocki2013,Witkowska2013,Lacki2017,Bialonczyk2018}. 
Subsequently, it has been recently shown that real-time evolution of time-independent many-body systems after a quantum quench, i.e. a sudden change of a control parameter of systems, across a critical value can reveal non-analytical behavior at certain moments of time \cite{Heyl2013,Hickey2014,Heyl2015,Budich2016,Sharma2016,Zunkovic2016,Bhattacharya2017a,Bhattacharya2017b,Karrasch2017,Heyl2017,Halimeh2017,Zauner-Stauber2017,Homrighausen2017,Lang2017,Weidinger2017,Chichinadze2017,Lerose2018,Mera2018,Piroli2018}. The critical behavior in time evolution has been also observed for quenches within the same phase   \cite{Andraschko2014,Vajna2014}. These phenomena are dubbed dynamical quantum phase transitions and they have been already demonstrated in experiments \cite{Jurcevic2017,Flaschner2018}, for review see \cite{Heyl2018_review}. 

In 2012 Frank Wilczek identified spontaneous breaking of continuous time translation symmetry into a discrete time translation symmetry by a quantum many-body system with the formation of a time crystal  \cite{Wilczek2012}. The original Wilczek's idea was proven to be impossible to realize, because he assumed a system in its ground state \cite{Bruno2013b,Watanabe2015,Syrwid2017,Iemini2017,Huang2017a,Prokofev2017}.  
 Nevertheless, soon the so-called discrete time crystals were proposed \cite{Sacha2015}, where periodically driven quantum many-body systems can spontaneously choose motion with a period different from the driving period.  Subsequent works \cite{Khemani16,ElseFTC,Yao2017,Lazarides2017,Russomanno2017,Zeng2017,Nakatsugawa2017,Ho2017,Huang2017,Gong2017,Wang2017,Tucker2018} eventually led to the experimental observation of the formation of this kind of crystalline structures in time in quantum many-body spin systems \cite{Zhang2017,Choi2017,Nayak2017,Pal2018,Rovny2018,Rovny2018a,Autti2018}.  (See also a recent work on experimental conditions needed for the realization of a time crystal with ultra-cold atoms bouncing on an oscillating mirror \cite{Giergiel2018}.)  It should be  mentioned that in the classical regime  spontaneous breaking of discrete time translation symmetry in atomic systems was also demonstrated in a laboratory \cite{Kim2006,Heo2010}.  

The new research area initiated by Frank Wilczek attracts substantial scientific attention. Recent progress in the field contains analysis of time quasicrystals \cite{Flicker2017,Giergiel2018discrete},  phase-space crystals \cite{Guo2013,Guo2016,Guo2016a,Liang2017},   topological time crystals \cite{Bomantara2018,Giergiel2018topological}  and analogs of condensed matter phenomena in the time domain \cite{Sacha15a,sacha16,Giergiel2017,Mierzejewski2017,delande17,Giergiel2017a,Mizuta2018}, for review see \cite{Sacha2017rev}. 
Also, it turns out that the concept of the dynamical quantum phase transitions can be extended to periodically driven quantum many-body systems \cite{Kosior2017}. That is, the return probability of a periodically evolving discrete time crystal reveals non-analytical behavior at a critical  moment of time after a quench to the non-interacting regime.  

In the present paper we return to the original Wilczek's idea \cite{Wilczek2012} of spontaneous breaking of continuous time translation symmetry. 
It has been shown that in the Wilczek model, spontaneous breaking of  continuous time translation symmetry can be observed if one restricts to specific excited eigenstates of the system \cite{Syrwid2017}. That is, the system prepared in a so-called yrast state is able to break spontaneously both space and time  translation symmetries and the localized center of mass of the many-body system reveals periodic motion which lasts forever if the number of particles $N\rightarrow\infty$.  


Here, we show that a quantum quench from the time crystal regime to the regime of the symmetric phase induces non-anlytical behavior of the return probability  of the system  to the initial state at the critical moments of time. In other words, we show that  the dynamical quantum phase transition occurs in systems with simultaneously broken continuous time and space translation symmetries. For other works on dynamical quantum phase transitions in systems with continuous symmetry breaking see \cite{Dora2013,Canovi2014,Fogarty2017,Weidinger2017}.  
The paper is organized as follows. In Sec.~\ref{timecrystal} we describe the Wilczek model and introduce the so-called continuum description of the system \cite{Zin2008a}. In Sec.~\ref{dynamical} the analysis of dynamical quantum phase transitions is presented and we conclude in Sec.~\ref{conclusion}.

\section{Time crystal}
\label{timecrystal}

Let us consider the Hamiltonian of $N$ bosons moving on a ring of a unit length,
\be
\hat H=\int\limits_0^1dx\;\hat\psi^\dagger\left(-\frac12\partial_x^2+\frac{g_0}{2}\hat\psi^\dagger\hat\psi\right)\hat\psi,
\label{h}
\ee
where we put $m=1$, $\hbar=1$ and where $\hat\psi(x)$ is the standard bosonic field operator and $g_0<0$ determines the strength of the attractive contact interactions between particles. 

In the mean field approach the ground state of the system is a Bose-Einstein condensate described by the product state $\psi_0(x_1,\dots,x_N)=\prod_{i=1}^N\phi_0(x_i)$ where $\phi_0$ is the lowest energy solution of the Gross-Pitaevskii equation \cite{Pethick2002},
\be
\left(-\frac12  \partial_x^2+g_0 (N-1) |\phi_0|^2\right)\phi_0=\mu\phi_0,
\label{gpe}
\ee
with $\mu$ being the chemical potential of the system. If the particles' interactions are very weak, then $\phi_0=1$ is the ground state solution of Eq.~(\ref{gpe}). However, when $g_0(N-1)<-\pi^2$, the mean field ground state becomes non-uniform and for $g_0(N-1)\rightarrow-\infty$ it is well approximated by the bright soliton solution  \cite{Carr2000}
\be
\phi_0(x)\propto \frac{1}{\cosh\left[g_0(N-1)(x-x_{\rm CM})/2\right]}.
\label{brsol}
\ee 
The position, $x_{\rm CM}$, of the center of mass of the system is arbitrary and it is determined in the process of spontaneous breaking of  continuous space translation symmetry \cite{Syrwid2017}. Indeed, the exact many-body ground state, which corresponds to the total momentum $P=0$, is also an eigenstate of the unitary operator which translates all particles by the same distance because such an operator commutes with the  $N$-body Hamiltonian (\ref{h}). However, the symmetric ground state is vulnerable to a perturbation and it is practically impossible to prepare it in an experiment if $N$ is large. Instead, experimentalists usually deal with the symmetry broken mean field solution $\phi_0$.

In 2012 Frank Wilczek proposed to introduce a magnetic flux through the ring along which particles are moving \cite{Wilczek2012}. Provided that the flux is appropriately chosen, he argued that a time crystal would form in the ground state.
In other words, spontaneous breaking of space translation symmetry and formation of a bright soliton would result also in the breaking of the time translation symmetry because the soliton was expected to move periodically along the ring.  Soon it has been shown that whatever the flux is chosen, the bright soliton will never  move if we consider the limit $N\rightarrow\infty$ \cite{Bruno2013b,Watanabe2015,Sacha2017rev}. However, the  spontaneous formation of the bright soliton which moves periodically along the ring does occur if the system is initially prepared in a yrast state, i.e. in the lowest energy eigenstate within the subspace corresponding to  the total momentum $P =2\pi N$ \cite{Syrwid2017}.  The liftetime of such a time crystal  goes to infinity with $N\rightarrow \infty$.

Thus, when we restrict to the Hilbert subspace with $P =2\pi N$, the critical value of the interaction strength, $g_0(N-1)=-\pi^2$, separates the regimes where both time and space translation symmetries are either preserved or can be spontaneously broken. In Sec.~\ref{dynamical} we show that a quantum quench from the symmetry broken regime  to the non-interacting regime results in the dynamical quantum phase transition that is observed in the non-analytical  evolution of the return probability of the system.
 If we start close to the critical value of the interaction strength, i.e. $g_0(N-1)\approx -\pi^2$, in order to obtain analytical predictions, we can apply the so-called continuum approximation \cite{Zin2008a}. Otherwise, we have to refer to numerical simulations or the mean field treatment. In the following when we consider a large number of particles we often assume that $g_0(N-1)\approx g_0N$.

For $g_0N\lesssim -\pi^2$, the eigenstate of the Hamiltonian (\ref{h}) corresponding to $P=2\pi N$ is not a Bose-Einstein condensate because the reduced single particle density matrix does not correspond to a pure state. However, only three single particle modes are substantially occupied by particles: the condensate mode corresponding to the momentum $k_0=2\pi$ and the two neighboring ones, i.e. $k_0\pm 2\pi$. Therefore, we can truncate the expansion of the bosonic field operator to three terms only, $\hat\psi(x)\approx e^{i k_0 x}\left( \hat a_0+e^{i2\pi x}\hat a_++e^{-i2\pi x}\hat a_- \right)$, where $\hat a$'s are the standard bosonic anihilation operators. In the limit of large $N$, $\hat a_0$ can be approximated by  $\sqrt{N-\hat a^\dagger_+\hat a_+-\hat a^\dagger_-\hat a_-}$ and the Hamiltonian (\ref{h}) reduces to \cite{Zin2008a}
\bea\label{H_approx}
\frac{\hat H}{2\pi^2}&\approx& \frac{\hat p_c^2+\hat p_s^2}{2}+V_{\rm eff}(\hat x_c,\hat x_s)+{\rm const.}, \\ && \cr
V_{\rm eff}&=&\frac{1+\alpha_0}{2}(\hat x_c^2+\hat x_s^2)-\frac{7\alpha_0}{32N}(\hat x_c^2+\hat x_s^2)^2,
\label{veff1}
\eea 
where 
\be\label{alfa}
\alpha_0=\frac{g_0N}{\pi^2},
\ee
and we have substituted 
\bea
\hat a_\pm&=&\frac12\left(\hat x_c  \pm i\hat x_s+i\hat p_c \mp  \hat p_s\right). 
\eea  
The conservation of the total momentum 
\bea
\hat P=2\pi\!\left(N \!+\! \hat a_+^\dagger\hat a_+ \!-\! \hat a_-^\dagger\hat a_-\right)=2\pi\!\left(N \!+\! \hat x_s\hat p_c \!-\! \hat x_c\hat p_s\right),
\eea
of the system is reflected by the rotational symmetry of the effective potential (\ref{veff1}).

Eventually, the initial many-body problem has been reduced to the Schr\"odinger equation of a fictitious particle in the two-dimensional space, 
\bea\label{schordinger_eff}
\frac{i}{2\pi^2}\partial_t\psi=-\frac12\nabla^2\psi+V_{\rm eff}(r)\psi,
\eea
where we have defined $\hat r^2=\hat x_c^2+\hat x_s^2$ \cite{Zin2008a}. For large $N$ we can approximate $\la \hat r^2 \ra /2 \simeq \la \hat n_c + \hat n_s\ra  \simeq  \la d\hat N\ra$ where \  $\la d\hat N\ra$ is the average number of particles depleted from the condensate \cite{Zin2008a}.  The wavefunction of a fictitious particle $\psi(r,t)$  provides information about the distribution of number of particles depleted from the condensate and we can interpret $p(r,t)\mbox{d}r$, where 
\be
p(r,t)= 2\pi r |\psi(r,t)|^2,
\label{pofr}
\ee  
is the probability of finding $2\, dN$ particles out of the condensate.

\section{Dynamical quantum phase transition}
\label{dynamical}

In this section we analyze  the dynamical quantum phase transition in the system of $N$ bosons with contact interactions \eqref{h}. Specifically,  we assume that the system is prepared in the lowest energy eigenstate within the $P=2\pi N$ subspace and consider a quench from the time crystal   regime to the non-interacting regime.

\subsection{Continuum approximation}

Let us start with the system in the vicinity of the critical point of the equilibrium quantum phase transition  $g_0N \lesssim-\pi^2 $. For large number of particles $N$, the many body Hamiltonian~\eqref{h} reduces to  the effective  Hamiltonian~\eqref{H_approx}, and the evolution of the system can be described by the effective 2D Schr\"{o}dinger equation \eqref{schordinger_eff} \cite{Zin2008a}. In the subspace with the total momentum $P=2\pi N$, the lowest energy state, aka the yrast state, can be well approximated by the harmonic oscillator ground state through the expansion of \eqref{veff1} around a local minimum 
\be\label{psi_0}
\psi(r,t=0) \propto e^{-(r-r_0)^2/(2b^2)},
\ee
where 
\bea
b&=&\left[-2(1+\alpha_0)\right]^{-1/4}, 
\label{bref}
\\ 
r_0 &=& \sqrt{\frac{8 N(\alpha_0 +1)}{7\alpha_0} } \equiv \sqrt{N} \;\tilde r_0,
\label{r0ref}
\eea
and $\alpha_0$, given by \eqref{alfa}, is the renormalized interaction strength.

The state \eqref{psi_0} is peaked around $r_0$, which accounts for low fluctuations of number of particles depleted from the condensate around the mean value $\la d\hat N\ra$. It is the eigenstate of the total momentum and consequently an eigenstate of the unitary operator which translates all particles by the same distance. Its time evolution is trivial and therefore it possesses also time translation symmetry. Since the interaction strength $g_0N $ exceeds the critical value $g_0N =-\pi^2 $, we expect that any small perturbation, such as the measurement of the position of one particle,  can lead to the breakdown of space and time translation symmetries, and consequently the  formation of a  time crystal \cite{Syrwid2017}.

Here, we do not consider the time crystal formation, but instead we choose \eqref{psi_0} as our initial state, and at $t=0$ we perform a quench to the noninteracting regime, $g_0N=0$.  For large $N$, the time evolution of \eqref{psi_0} can be found analytically (see the Appendix) 
\be
\psi(r,t)\propto  e^{-i \frac{r^2}{2} \cot(2\pi^2 t)}\left(   h^{(+)}(r,t) +  h^{(-)}(r,t) \right),
\ee
where 
\bea
h^{(\pm)}(r,t) &=& e^{\mp i \pi/4} \frac{H_{-3/2}\left(-\frac{\gamma^{(\pm)}(r,t)}{2 \sqrt{\alpha(t)}}\right)}{\sqrt{r \sin( 2\pi^2 t) \alpha(t)^{3/2}}}, \\ 
\gamma^{(\pm)}(r,t) &=& \frac{r_0}{b^2} \pm i \,\frac{r}{\sin(2\pi^2 t)}, \\
\alpha(t) &=& \frac{1}{2}\left(b^{-2} -  i\cot(2\pi^2 t ) \right),
\eea
and $H_{-3/2}(z)$ is a Hermite function of degree $-3/2$ \cite{Lebedev1965}.

\begin{figure}[tb] 	          
\includegraphics[width=1.\columnwidth]{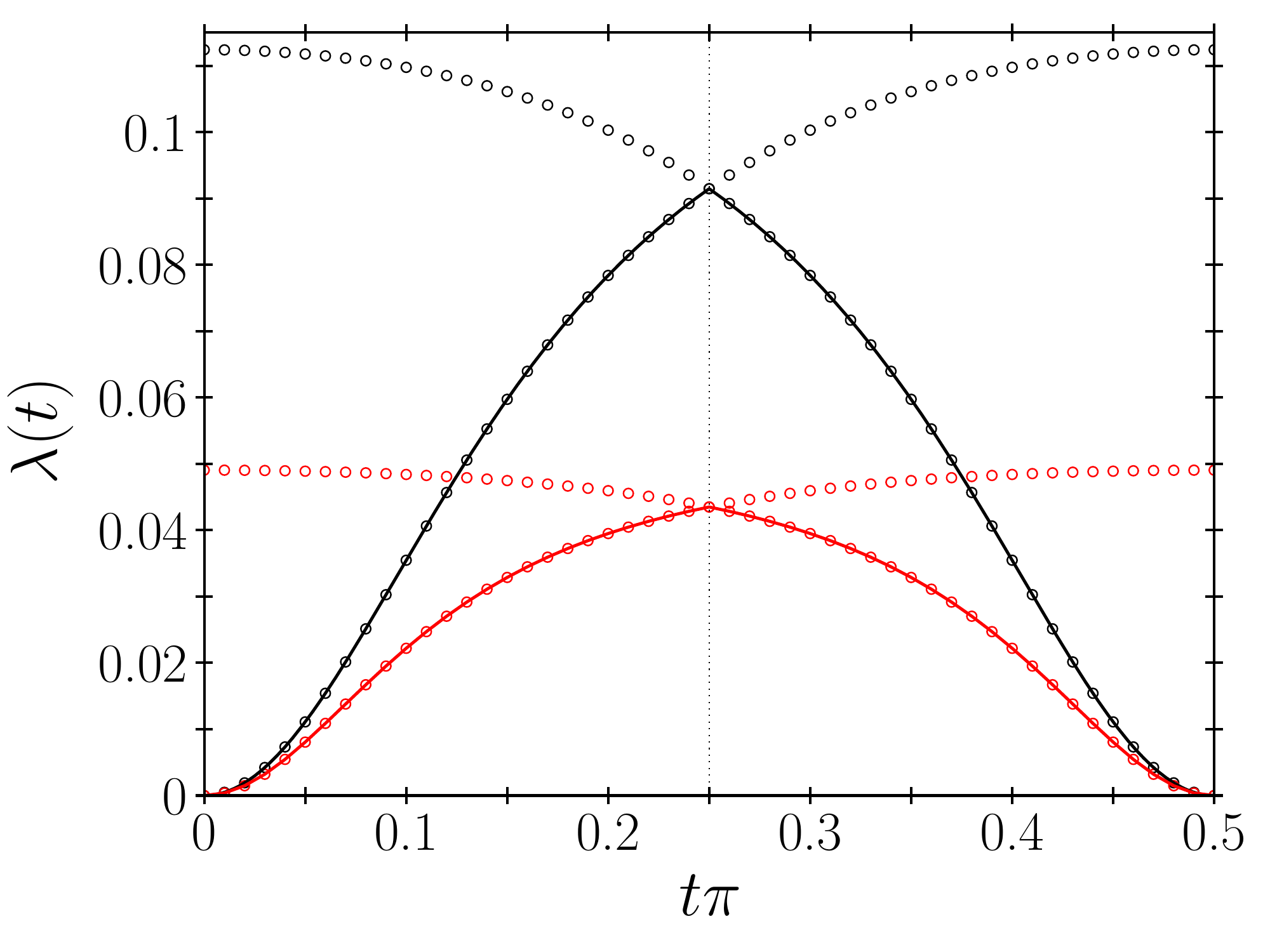}    
\caption{The rate function of the Loschmidt echo for two values of the initial interaction strength $g_0N = - 10.5$ (red solid) and $g_0N =-11$ (black solid). The rate function can be approximated by $\lambda(t)\approx \min[\lambda_+(t),\lambda_-(t)] $, where $\lambda_\pm(t)$ (red and black circles) are auxiliary functions, Eq.~\eqref{lambda_pm}. Since the rate $\lambda(t)$ is  given by a minimum of $\lambda_\pm(t)$, it is not differentiable at their crossing at the critical moment of time $t_c=1/4\pi$, indicated by vertical dotted line, which can be associate with a dynamical quantum phase transition. All units are dimensionless.
}
\label{rates}   
\end{figure}

Following Heyl et.al~\cite{Heyl2013} we associate the dynamical quantum phase transition with the non-analyticity of the Loschmidt echo
\be\label{l_echo}
\mathcal L(t) = \left|  \la\psi(0) | \psi(t) \ra\right|^2,
\ee 
i.e., the return probability of  $|\psi(t)\ra$ to the initial state $|\psi(0)\ra$ after the quench. As the Loschmidt echo decays exponentially with the system size \cite{Heyl2018_review}, it is convenient to analyze the rate function
\be\label{rate_def}
\lambda(t) \equiv -\lim_{N\rightarrow \infty} \lambda^{(N)}(t)  = -\lim_{N\rightarrow \infty}\frac{1}{N} \ln \mathcal L(t),
\ee
which can be measured in experiments  \cite{Jurcevic2017,Kosior2017}.  Within the three mode approximation near the equilibrium critical point we find that (see Appendix)
\be\label{lambdat}
\lambda(t) \approx \min[\lambda_+(t),\lambda_-(t)], 
\ee
where 
\bea
\lambda_+(t) &=& \frac{2 \tilde r_0^2 b^2}{\left[ b^4 +\tan^2(\pi^2 t)\right]}, \nonumber \\
\lambda_-(t) &=& \frac{2 \tilde r_0^2 b^2}{ \left[ b^4 +\cot^2(\pi^2 t)\right]}, \label{lambda_pm}
\eea
and $b$ and $\tilde r_0$ are given in (\ref{bref})-(\ref{r0ref}). Since the rate function in the limit $N\rightarrow\infty$ is given by a minimum of two functions $\lambda_\pm(t)$, it has the non-analytic cusps at the critial time $t_c=1/4\pi$ when $\lambda_+(t_c)=\lambda_-(t_c)$, which is illustrated in Fig.~\ref{rates}. The non-analyticity of the rate function  results in the discontinuity of the first time derivative and appear at $t=t_c=1/4\pi$
\be
 \left|\dot \lambda_+(t_c) -\dot \lambda_-(t_c)\right|  =  \frac{16 \pi^2 \tilde r_0^2   b^2}{  (1+b^4)^2 }. 
\ee

\begin{figure}[tb]
\includegraphics[width=1\columnwidth]{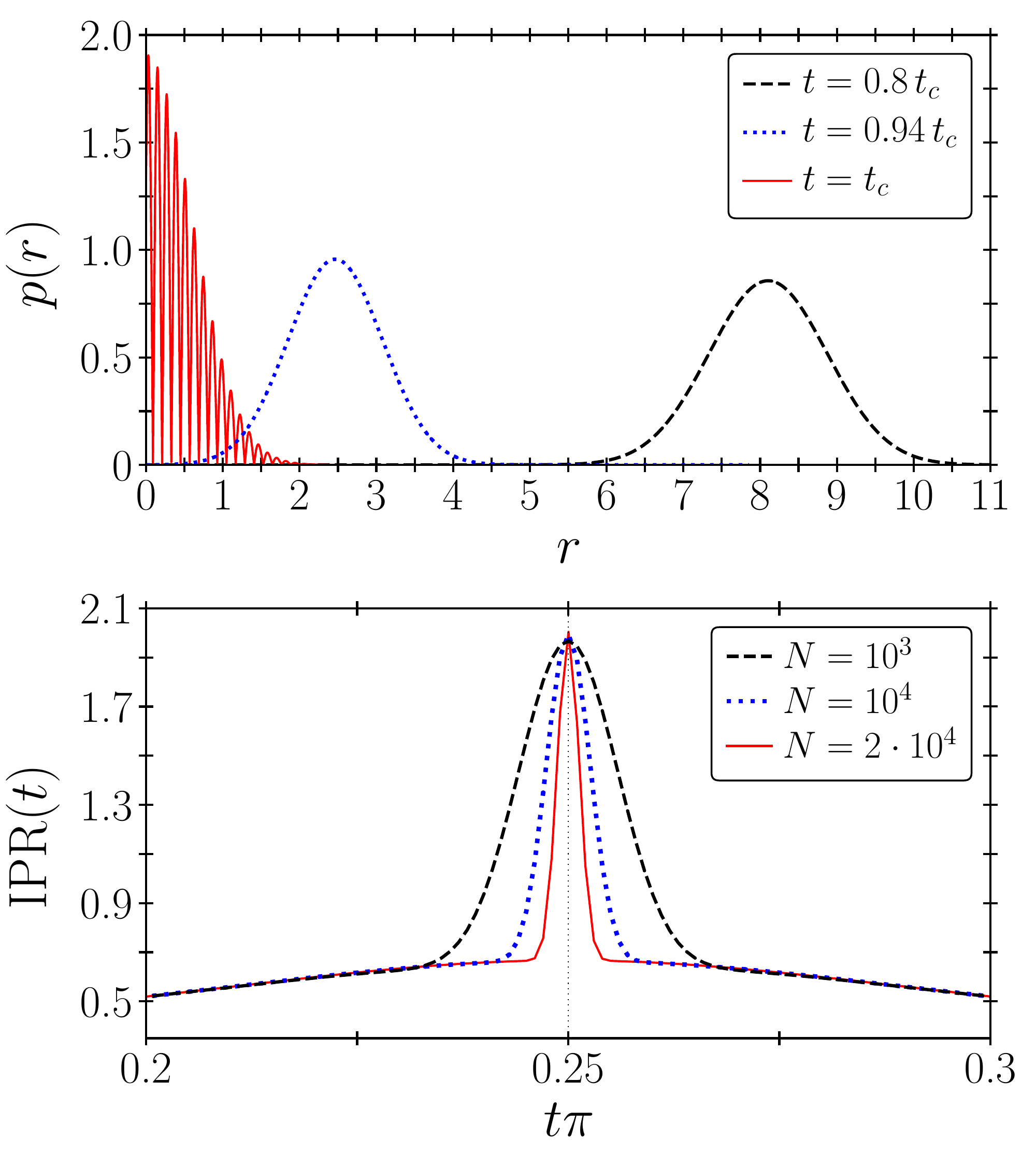} 
 \caption{Increase of the purity of the many-body state at the critical time $t_c$.  Top panel: the probability density $p(r,t)=2\pi r |\psi(r,t)|^2$  at three different moments of time $t=0.8\,t_c$ (dashed black), $t=0.94\,t_c$ (dotted blue) and $t=t_c$ (solid red) for $N=10^4$. The interference fringes visible in the plot corresponding to $t=t_c$ result in the decrease of uncertainty of the number of particles depleted from the condensate, hence, the increase of the purity of the many-body state.  Bottom panel: the normalized inverse participation ratio (IPR) near the critical time $t_c=1/4\pi$, indicated by vertical dotted line,
 for three different total numbers of particles
  $N=10^3$ (dashed black), $N=5\cdot 10^3$ (dotted blue) and $N=2\cdot 10^4$ (solid red). The change of IPR becomes more rapid with increasing $N$. All units are dimensionless.}
 \label{prob_ipr}  
\end{figure} 

In  Ref.~\cite{Kosior2017} dynamical quantum phase transitions in discrete time crystals have been investigated. There, it has been pointed out that not only the Loschmidt echo has non-analytic points, but also the von Neuman entropy of the reduced density matrix is discontinuous in the thermodynamic limit at the same critial time~$t_c$. The latter can be identified with the momentary increase of purity of the many-body quantum state.  Here, we observe a very similar behavior.  In Fig.~\ref{prob_ipr}~(top panel) we plot the probability density $p(r,t)$ (\ref{pofr}) at three different moments of time 
$t=0.8\,t_c$, $t=0.94\,t_c$ and $t=t_c$. In the course of time evolution the probability density $p(r,t)$ can be well approximated by a normal distribution with  an almost constant variance and the mean value which approaches the center of the effective potential. Eventually, at $t= t_c$ the fictitious particle  reaches the center which accounts for the interference fringes  visible in the plot of $p(r,t_c)$. 
Following the discussion in Sec.~\ref{timecrystal}, we can interpret $p(r,t)$ as the probability distribution of number of particles depleted from the condensate. Hence, the interference fringes  in Fig.~\ref{prob_ipr}~(top panel) are related to a drastic reduction of uncertainty of the number of particles depleted from the condensate what reflects an increase of purity of the many-body state.
As a measure of the purity we investigate the normalized inverse participation ratio (IPR), i.e.,
\be
\mbox{IPR} (t) = \int\mbox{d}r \,p(r,t)^2 = \int\mbox{d}r \, r^2 |\psi(r,t)|^4,
\ee
which is a measure of localization of states: larger values of IPR correspond to more localized states.
 In Fig.~\ref{prob_ipr}~(bottom panel) we plot IPR in the vicinity of the critical time $t_c$ for three different total number of particles $N=10^3$, $N=5\cdot 10^3$ and $N=2\cdot 10^4$. Indeed, IPR is sharply peaked around $t=t_c$ and the peak is the narrower, the larger $N$ is.

\subsection{Numerical results}

The continuum approximation  
allows us to study the case of interparticle interactions that are very close to the critical point only, i.e., $g_0(N-1)\approx -\pi^2$. In order to examine the Loschmidt echo further away from the critical point we employ the numerical simulations of the many-body system (\ref{h}) in a truncated Hilbert space. The yrast state corresponding to the total momentum $P=2\pi N$ can be represented in the Fock state basis $\left|\{n\}\right>=\prod_{j=-J}^{J}\left|n_j\right>$, where $n_j$  denotes the number of particles occupying a single particle state $\varphi_j(x)=\mathrm{exp}\left[i 2\pi (j+1) x\right]$, and the parameter $J$ has to be chosen sufficiently big in order to achieve converged results. 

\begin{figure}[tb] 	            
\includegraphics[width=1\columnwidth]{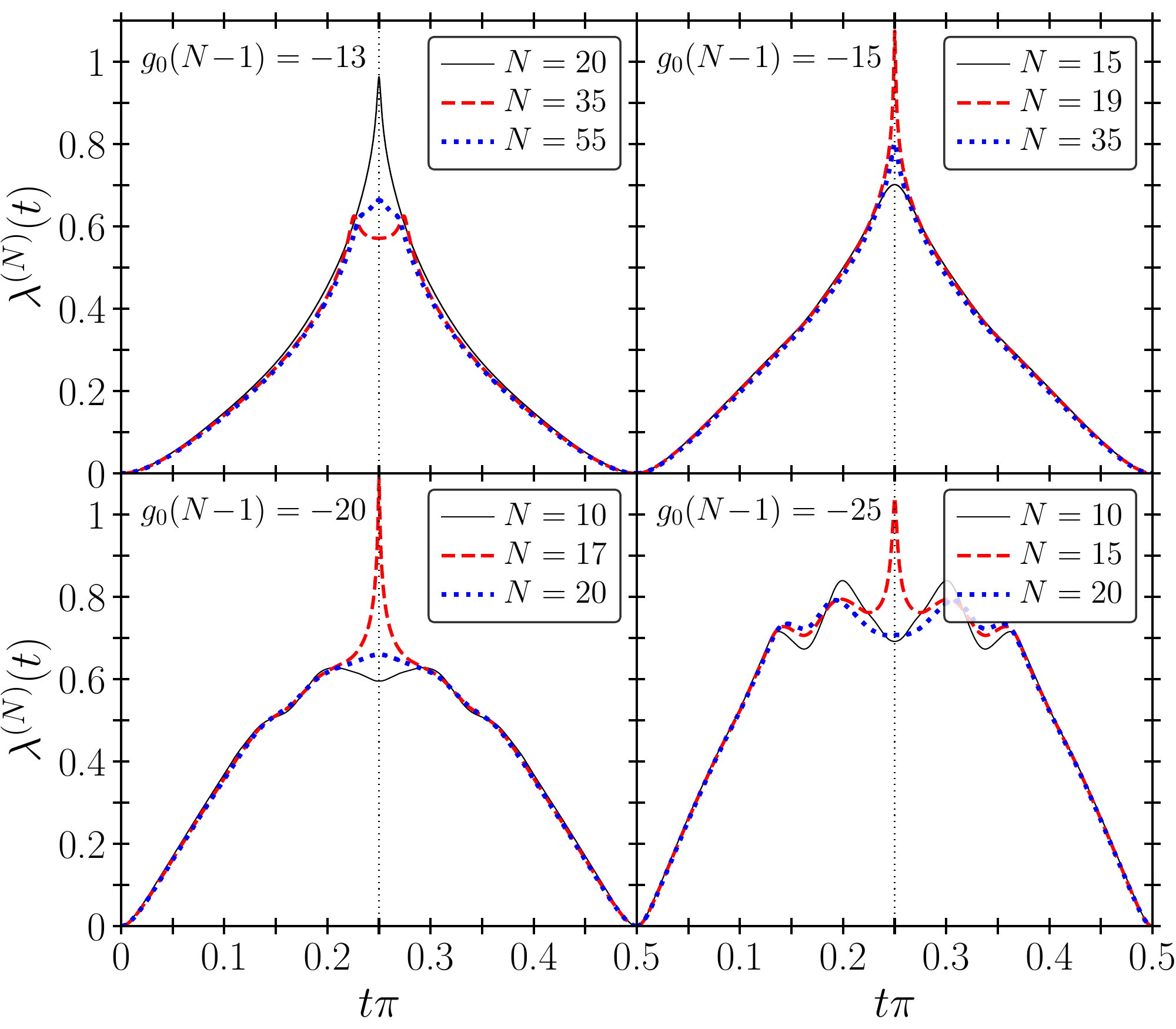}  
\caption{The rate function $\lambda^{(N)}(t)$  for different initial values of inter-particle attractions far from the critical point, i.e. from left top to right bottom $g_0(N-1)=-13,-15,-20,-25$.  In each panel we compare the results obtained for different number of particles $N$. Note that the rates  $\lambda^{(N)}(t)$ can  change quite  quickly with $N$ in the vicinity of the anticipated critical time $t_c=1/4\pi$ that is marked with vertical dotted line. The rates tend to diverge for some specific values of $N$, which has been also observed in other systems \cite{Heyl2018_review,Kosior2017}. All units are dimensionless.
}
\label{mb_rates}   
\end{figure}

Similarly as in the case of the continuum approximation we are interested in a behavior of the Loschmidt echo $\mathcal{L}(t)$ after the quench from the time crystal regime to the non-interacting regime if the many-body system is initially prepared in the yrast state with $P=2\pi N$.
Although, within the numerical simulations the thermodynamic limit is not attainable, we show that the rates $\lambda^{(N)}(t)=-N^{-1}\ln \mathcal{L}(t)$ obtained for finite numbers of particles $N$ resemble the cusp-like non-analytic behavior, see Fig.~\ref{mb_rates} and the discussion in Sec.~\ref{SB}. Note that  while the tails of the rates are almost insensitive to  a change of the total particle number, the shape of $\lambda^{(N)}(t)$ can vary quite quickly in the vicinity of the critical point.  Moreover, the rates tend to diverge for some specific values of $N$. The latter phenomenon has been also observed in other systems \cite{Heyl2018_review,Kosior2017}.

\subsection{Symmetry broken state}\label{SB}

So far we have analyzed the return probability of the system after the quench from the time crystal regime to the non-interacting regime starting with the yrast state, i.e. with a state which preserves both time and space translation symmetries.
However, such an yrast state is very fragile and any perturbation of the system can lead to spontaneous breaking of the  translation symmetries and  to the time crystal  formation \cite{Syrwid2017}. 

Let us assume now that the system containing a large number of particles is initially  prepared in the symmetry broken state, i.e., $\psi  =\Pi_{i=1}^N\phi(x_i,t)$, where $\phi(x,t)$ is  the mean-field bright soliton moving periodically on a ring of a unit length with the momentum $P/N= 2\pi$. The single-particle  wave-function $\phi(x,t)$  is a solution of the Gross-Pitaevskii equation and corresponds to the lowest energy solution in the frame rotating with the frequency $2\pi$. For sufficiently strong interactions
the desired single-particle wave-function in the rotating frame may be approximated by  $\tilde \phi_{x_{\rm CM}}(x) =  e^{i 2  \pi x} \phi_0 (x)\propto e^{i 2  \pi x}/\cosh[g_0N(x-x_{\rm CM})/2]$.   Such a  soliton  is parametrized by  the center of mass position $x_{\rm CM}$  which can be arbitrary and which is determined in a spontaneous symmetry breaking process. 

 Similarly as in the previous subsections we are interested in the non-equilibrium dynamics of the system after the quench, i.e. when  $g_0N$  is suddenly set to zero.  The initial symmetry broken state $\tilde\psi=\Pi_{i=1}^N\tilde\phi_{x_{\rm CM}}(x_i)$ belongs to the degenerate subspace parametrized by the continuous parameter $x_{\rm CM}$. We are interested in the return probability of $\tilde\psi(t)$ to the degenerate ground state manifold after the quench, i.e. when at $t=0$ the interactions between particles are turned off.  A natural generalization of the Loschmidt echo for the continuous symmetry broken solutions is the following \cite{Weidinger2017}
\be\label{l_echo_mf}
\mathcal{L}_{\rm SB} (t)  = \int \mbox{d}x_{\rm CM} \left| \la \tilde \phi_{x_{\rm CM}} | \tilde \phi (t)  \ra \right|^{2N} \propto e^{-N\lambda_{\rm SB}(t)}, 
\ee
where $ \tilde \phi (t) $   evolves according to the Gross-Pitaevskii equation with $g_0=0$. For simplicity we choose the initial state $\tilde \phi(t=0)=\tilde \phi_{x_{\rm CM}}$ to be localized around $x_{\rm CM}=0$.
\begin{figure}[tb] 	         
\includegraphics[width=1\columnwidth]{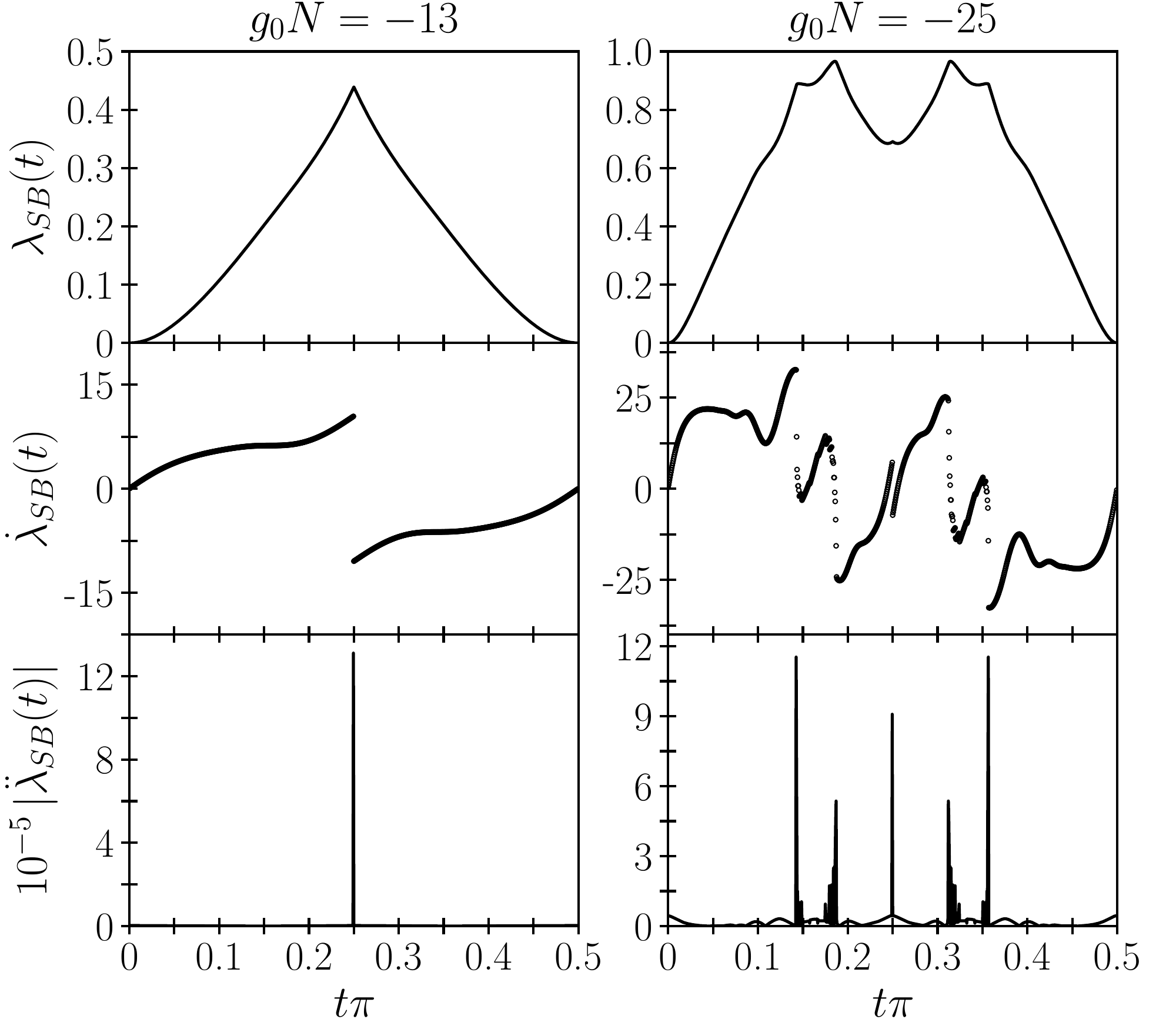}   
 \caption{ The plots of the rate function $\lambda_{\rm SB}(t)$,  its first derivate $\dot \lambda_{\rm SB}(t)$ and the absolute value of the second $|\ddot\lambda_{\rm SB}(t)|$  for two interactions strengths $g_0N =-13$  and $g_0N =-25$ (left and right column respectively). The singularity at critical moments of time are distinguished with a cusp of $\lambda_{\rm SB}(t)$, discontinuity of $\dot \lambda(t) $ and a delta-kick of the second derivate $\ddot \lambda_{\rm SB}(t)$.  For $g_0N =-13$,  we obtain a single critical time $t_c=1/4\pi$, whilst for $g_0N =-25$ we observe a cascade of singular points, see the discussion in the main text. The presented mean-field results correspond to $N\rightarrow\infty$. Many-body results for the system prepared in the symmetry-preserving state but for a finite $N$ are shown in Fig.~\ref{mb_rates}. All units are dimensionless.
}
\label{rates_mf}   
\end{figure} 

In the case of a system with $M$-fold degeneracy of the ground state level, the Loschmidt echo \eqref{l_echo} for the symmetric ground state is the same as the generalized Lochmidt echo \eqref{l_echo_mf} defined for a symmetry broken state if the thermodynamic limit is considered.
 Here,  we deal with the continuous symmetry breaking. Since the symmetry broken solitonic solutions, corresponding to different values of $x_{\rm CM}$ are not mutually orthogonal  and forms an overcomplete  basis, the  two defintions \eqref{l_echo} and \eqref{l_echo_mf}  might not be equivalent.  In the following we point out that although there are quantitative differences between the results of \eqref{l_echo} and \eqref{l_echo_mf}, both of them reveal non-analytical behavior around critical moments of time.

\begin{figure}[tb] 	       
\includegraphics[width=1\columnwidth]{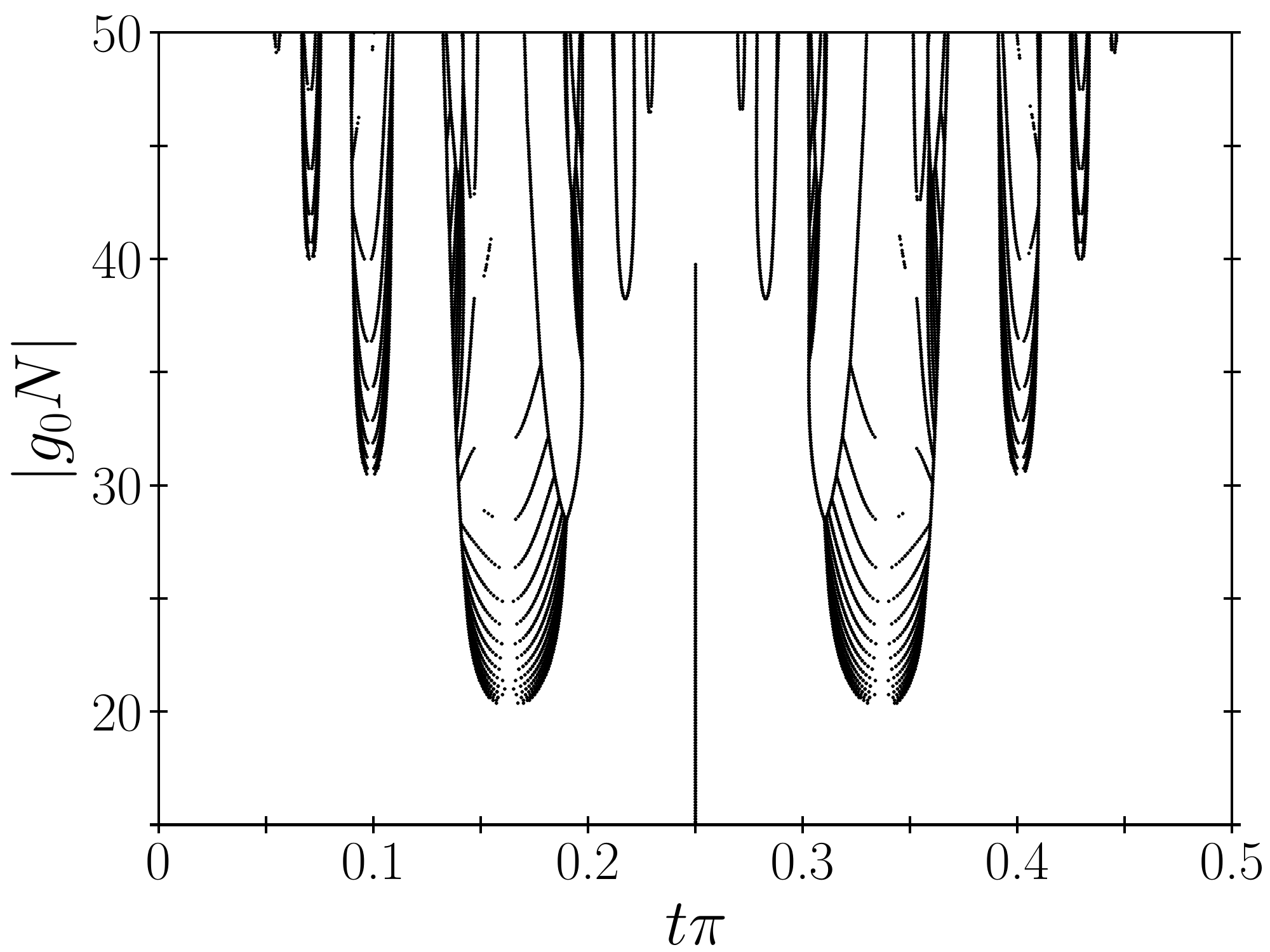}       
\caption{
Singular points of the absolute value of the second derivative of the rate function  \eqref{rate_mf}, i.e., $|\ddot \lambda_{\rm SB} (t)|$, in the space spanned by the interaction strength $g_0N$ and time $t$. All units are dimensionless.
}
\label{cascades}   
\end{figure}

Let us define
$
\lambda_{x_{\rm CM}}(t) = -\ln \big| \la \tilde \phi_{x_{\rm CM}} | \tilde \phi (t)  \ra \big|^2
$
 and estimate the rate function  \eqref{l_echo_mf}
using the steepest descent method
\bea
\lambda_{\rm SB}(t)
&=&  -\lim_{N\rightarrow\infty}\frac{1}{N}\ln \int \mbox{d}x_{\rm CM} e^{ -N \lambda_{x_{\rm CM}}(t)} \nonumber \\ && \cr 
&\approx& 
-\lim_{N\rightarrow\infty}\frac{1}{N}\ln \left\{\exp\left[ - N \min_{x_{\rm CM}\in [0,1)}\lambda_{x_{\rm CM}}(t) \right] \right\} \nonumber \\ && \cr
&=&  \min_{x_{\rm CM}\in [0,1/2]}\lambda_{x_{\rm CM}}(t) \label{rate_mf} . 
\eea
In Fig.~\ref{rates_mf} we present plots  of: $\lambda_{\rm SB}(t)$,  its first time derivate $\dot \lambda_{\rm SB}(t)$ and the absolute value of the second derivate $|\ddot\lambda_{\rm SB}(t)|$  for two different interaction strength $g_0N =-13$  and $g_0N =-25$. The rates are periodic with the period $T=1/ 2\pi$ which   corresponds to the half of the revival time  for non-intercating particles on the ring.  Within one period, for $g_0N =-13$, we obtain a single critial time $t_c=1/4\pi $ which  coincides with $t_c$ observed in  the previous   subsections.  On the other hand, for $g_0N =-25$  there is  a sequence of singularities. 
 The presence of the sequence of singularities is specific to the ring geometry. After the quench to the non-interacting regime, the evolution of $\lambda_{x_{\rm CM}}(t)$ can be easily calculated in the momentum space, 
\bea
 \lambda_{x_{\rm CM}}(t)&=&-\ln\left| \la \tilde \phi_{x_{\rm CM}} | \tilde \phi (t)  \ra \right|^2  
 \cr &=&   
 -\ln\left| \sum_{k_n} |\phi_0(k_n)|^2 e^{-ik_n^2t/2 +i k_n x_{\rm CM} } \right|^2, 
\eea
where $k_n=2\pi n$ and $\phi_0(k)$ is the Fourier transform of a solitonic solution $\phi_0(x)$ of the Gross-Pitaevskii equation which for $g_0N\ll -\pi^2$ is given by (\ref{brsol}).
For $g_0N\lesssim -\pi^2$, the translational symmetry is merely broken, therefore  $\phi_0(x) \propto 1+ \delta \cos \left(2 \pi x \right)$ where $\delta \ll 1$. Consequently, for $g_0N \lesssim -\pi^2$ we obtain 
\begin{gather}
 \lambda_{x_{\rm CM}}(t) \simeq 
   -\ln \left[1+  4\delta^2 \cos(2\pi^2 t) \cos(2\pi x_{\rm CM}) \right] + {\rm const.}, \label{rate_approx} 
\end{gather}
which is minimized by $ x_{\rm CM} =0 $ when $t<1/4\pi$ or by $ x_{\rm CM} =1/2 $ when $t>1/4\pi$.
At $t=t_c$ the rate 
$\lambda_{\rm SB}(t)$ given by \eqref{rate_mf} is not differentiable what is associated with the dynamical quantum phase transition. The approximate formula \eqref{rate_approx} breaks down when we increase the interaction strength because a larger number of momentum modes is required to describe the evolution of $\lambda_{x_{\rm CM}}(t)$ if the soliton is initially strongly localized. Consequently different rates $\lambda_{x_{\rm CM}}(t)$ corresponding to different $x_{\rm CM}$ become minimal for different moments of time and the sequence of critical moments of time, visible in Fig.~\ref{rates_mf}, turns up. For $g_0N\lesssim -20$, there is a single critical time at $t_c= 1/4\pi$ which disappears for $g_0N\approx -40$, see Fig.~\ref{cascades}. For $g_0N\gtrsim -20$ a virtual cascade of singular points in time develops due to the narrowing of the solitonic solution, cf. Fig.~\ref{cascades}.

If we start with the symmetry broken state but close to the critical point, we also observe a non-analytical behavior in time but the obtained values of $\lambda_{\rm SB}(t)$ (\ref{rate_mf}) do not precisely match $\lambda(t)$ (\ref{lambdat}).

\section{Summary}
\label{conclusion}

We have analyzed bosons on a ring with attractive contact interactions, i.e., a many-body system which is able to break spontaneously continuous time and space translation symmetries. The system is related to the Wilczek model of a time crystal, but contrary to the original Wilczek idea, we do not consider the ground state of the system but the lowest energy eigenstate within a subspace of a non-zero total momentum, the so-called yrast state. 

When the attractive interactions between bosons are sufficiently strong, the mean-field description predicts formation of a bright soliton which can move periodically on the ring. In the full many-body approach, the corresponding many-body eigenstate is the yrast state with a non-zero total momentum. The many-body eigenstate fulfills continuous time and space translation symmetries but in the limit when $N\rightarrow\infty$ it is extremely fragile and breaks the symmetries under an infinitesimally weak perturbation and thus spontaneously. The symmetry breaking leads to the formation of a bright soliton which is moving periodically on the ring, i.e a time crystal emerges.

We have analyzed dynamical quantum phase transitions in the system when the interactions between particles are suddenly turned off. Such a quantum quench from the time crystal regime to the non-interacting regime results in a non-analytical behavior of the return probability of the evolving state to the initial yrast state. Starting close to the critical value of the interaction strength for the bright soliton formation, an analytical description of the dynamical quantum phase transition has been carried out. Away from the critical point, signatures of the dynamical quantum phase transitions have been analyzed with the help of numerical simulations. We have also investigated the quench to the non-interacting regime when the system is initially prepared in the mean-field bright soliton state which breaks time and space translation symmetries. The generalized return probability reveals also a non-analytical behavior which becomes the more complex, the stronger particle interactions are. 

To conclude the considered system is an experimentally attainable system where time and space translation symmetries can be spontaneously broken and dynamical quantum phase transitions can be observed.

\section*{Acknowledgement}

Support of the National Science Centre, Poland via Projects No.~2016/21/B/ST2/01086 (A.K.), No.~2016/21/B/ST2/01095 (A.S.) and under QuantERA programme No.~2017/25/Z/ST2/03027 (K.S.) is acknowledged.

\section*{APPENDIX}
\renewcommand{\theequation}{A.\arabic{equation}}
\setcounter{equation}{0}

In a vicinity of the quantum critical point  $\alpha_0 =g_0N / \pi^2 \lesssim -1$, within the continuum approximation the yrast state with the total momentum $P=2\pi N$  is described by a wavefunction of a fictitious particle \eqref{schordinger_eff}

\be\label{app:psi_0}
\psi(r,t=0) = \mathcal N e^{-(r-r_0)^2/(2b^2)}.
\ee
where 
$\mathcal N$ is the normalization constant and
\bea
b&=&\frac{1}{\left[-2(1+\alpha_0)\right]^{1/4}},  \\
r_0 &=& \sqrt{\frac{8 N(\alpha_0 +1)}{7\alpha_0} } \equiv \sqrt{N} \tilde r_0. 
\eea   

After a quench across a quantum critical point to $\alpha_0=0$, i.e. to the noninteracting systems, the effective potential \eqref{veff1} is simply a  harmonic potential.  Hence, the evolution of \eqref{app:psi_0} is given by
\be\label{app:psit_from_prop}
\psi(r,t)= \int_0^{2\pi}  \mbox{d}\phi' \int_0^\infty\mbox{d}r' \,r' K(r,\phi,t;r',\phi') \psi_0(r'),
\ee
where 
\be
 K(r,\phi,t;r',\phi')  = \frac{i \exp\left( 
\frac{i \cot(\omega t)(r^2+r'^2) }{2  t}  - \frac{ r r' \cos(\phi -\phi')}{ \sin (\omega t) } \right)}{2 \pi \sin (\omega t)} 
\ee
is a 2D harmonic oscillator propagator in the polar coordinates, and $\omega =2 \pi^2$. The integration over polar angle 
is straightforward, as it gives the integral representation of the zeroth order Bessel function of the first kind $\mathcal J_0 (x)$
\be
\int_0^{2\pi}\mbox{d} \phi' \, e^{-i x \cos ( \phi - \phi') } \equiv  2 \pi \mathcal J_0 (x).
\ee
Consequently, we can write
\bea
\psi (r,t) &=& \mathcal N (t) \chi (r,t) e^{- r_0^2/2b^2}e^{i\cot(\omega t) r^2/2}, \\
\label{app:chi}
\chi (r,t) & = & \int_0^\infty \mbox{d}r'r' \mathcal J_0 \left( \frac{r r'}{\sin(\omega t)} \right)e^{  \alpha(t)r'^2} e^{ \beta r'},
\eea
where 
\be
\alpha(t) = 1/2b^2 -i \cot (\omega t) /2
\ee
and $\mathcal N(t) = -i \,\mathcal N / \sin (\omega t) $, $\beta = r_0/b^2$. 
The integration over the radial part of \eqref{app:chi} can be performed efficiently by approximating the Bessel function with its asymptotic form $
\mathcal J_0 (x) \approx \sqrt{\frac{2}{\pi x}} \cos (x -\frac{\pi}{4} )$,
which yields

\begin{widetext}
\begin{multline}\label{app:chi_approx}
\chi (r,t) \approx \sqrt{\frac{ \sin (\omega t)}{2\pi r}} \int_0^\infty \mbox{d}r' \sqrt{r} e^{-\alpha(t) r'^2}e^{\beta r'}   \left(
e^{-i \frac{\pi}{4}  + \frac{i r r'}{\sin (\omega t)} } +
 e^{+i \frac{\pi}{4}  - \frac{i r r'}{\sin (\omega t)} } 
\right) = \\
= \sqrt{\frac{\sin (\omega t)}{4 \,r \,\alpha(t)^{3/2}}}\left[
 e^{-i\frac{\pi}{4} }
H_{-3/2} \left( \frac{-\gamma_+(r,t)}{2\sqrt{\alpha(t)}} \right) 
+ e^{+i\frac{\pi}{4} }
H_{-3/2} \left( \frac{-\gamma_-(r,t)}{2\sqrt{\alpha(t)}} \right)
\right], 
\end{multline}
  \end{widetext}
where 
\be\label{app:gamma}
\gamma^{(\pm)}(r,t) = r_0/b^2 \pm i \,r/\sin(2\pi^2 t)
\ee
and $H_{\nu}(z)$ is a Hermite function of degree $\nu$ \cite{Lebedev1965}. In  \eqref{app:chi_approx} we have employed the integral
\cite{Ryzhik2015}
\be
\int_0^\infty \mbox{d}x x^{\nu -1} e^{-\beta x^2 -\gamma x}=
\frac{\Gamma(\nu)}{(2\beta)^{\frac{\nu}{2}}}e^{\frac{\gamma^2}{8\beta}} D_{-\nu}\left(\frac{\gamma}{\sqrt{2\beta}}\right)
\ee
where $\Gamma(z)$ is the Euler gamma and $D_\nu(z)$ is the parabolic cylinder function, and we have made use of the identity \cite{Lebedev1965}
\be
D_\nu(z) =  2^{-\frac{\nu}{2}} e^{-\frac{z^2}{4}} H_\nu\left(\frac{z}{\sqrt{2}}\right).
\ee 

Now, we are ready to calculate the Loschmidt echo between the initial yrast state $\psi(r,t=0)$ \eqref{app:psi_0} and the time evolved state $\psi(r,t)$  after the quench at $t=0$. Let us write

\be\label{app:echo}
\mathcal L(t) = \left|  \la \psi(0) | \psi(t) \ra\right|^2 = \frac{4\pi^2 \mathcal{N} ^4}{\sin^2 (\omega t)} e^{-2r_0^2/b^2} |I(t)|^2, 
\ee 
where
\be\label{app:It}
I(t) =  \int_0^\infty \mbox{d}r\, r e^{ -\alpha(t)r^2+\beta r}\chi(r,t).
\ee
Remembering that in the end we are interested in the rate $\lambda(t) = -\lim_{N\rightarrow\infty} N^{-1} \ln\mathcal  L(t)$,  we can approximate $\chi(t)$ in \eqref{app:It} in the limit of large number of particles $N$. Since for $\omega t \ne 0 \,(\mbox{mod }\pi)$ the argument of the Hermite function in \eqref{app:chi} is proportional to $\sqrt{N}$, we can apply an asymptotic representation of the Hermite functions \cite{Lebedev1965}
\be
H_\nu(z) \approx \frac{\sqrt{\pi}}{\Gamma(-\nu)} e^{z^2}z^{-\nu-1},
\ee
where we have dropped an irrelevant phase factor. Therefore, for sufficiently large $N$ we can write
\begin{gather}
|I(t)|^2 \approx \frac{|\sin(\omega t)|}{2 |\alpha(t)|^2} \left|I_+(t)+I_-(t)\right|^2 \\
\approx 
 \frac{|\sin(\omega t)|}{2 |\alpha(t)|^2}  \max\left(|I_+(t)|^2,|I_-(t)|^2\right) \label{app:It_approx}
,
\end{gather}
where
\bea
|I_\pm(t)| &=& \left| \int_0^\infty \mbox{d}r \sqrt{-r\gamma_\pm(r,t)} e^{\Lambda_\pm(r,t)}   \right|,  \nonumber \\
\Lambda_\pm(r,t) &=& -\alpha(t) r^2 +\beta r +\frac{\gamma_\pm^2(r,t)}{4\alpha(t)}.
\eea
As we shall see,  $|I_\pm(t)|$ drop exponentially with $N$, therefore the approximation made in \eqref{app:It_approx} is justified. 
Since the the real part of $\alpha(t) \propto 1  $ is positive and the argument minimizing  $\mbox{Re} [ \Lambda_\pm(r,t)]$, i.e.
\bea
r_\pm^{min}(t) &=& r_0 \frac{\xi(t) \mp \cos(\omega t)}{\xi(t)+1},  \\
 \nonumber
 \xi(t) &=&  \sin^2(\omega t) \left( 1/b^4 + \cot^2(\omega t) \right),
\eea
 is proportional to $r_0 \propto \sqrt{N}$, we can approximate 
 \begin{gather}
|I_\pm(t)| \approx \left| \sqrt{-r_\pm^{min}(t)\gamma_\pm\left[r_\pm^{min}(t),t\right]} \int_{-\infty}^\infty \mbox{d}r  e^{\Lambda_\pm(r,t)}   \right| = \nonumber \\ 
= \left| \frac{  \sqrt{-r_\pm^{min}(t)\gamma_\pm\left[r_\pm^{min}(t),t\right]} }{c(t)} e^{\zeta_\pm(t)}   \right| ,
 \end{gather}
where
\be
c(t)  =  \frac{1}{2\pi} \frac{1 + b^4- 2 i b^{2} \cot( 2\omega t) }{ b^{2}  -  i  b^4  \cot (2\omega t)},
\ee
and
\bea
\zeta_+(t) &=& \frac{1}{b^2} \frac{r_0^2}{1 -i b^2 \cot( \omega t)  }  , \nonumber \\ 
\zeta_-(t) &=& \frac{1}{b^2}  \frac{r_0^2}{1 +i b^2 \tan( \omega t)  } .   
\eea
Finally, after a straightforward calculation one gets
\begin{gather}
\lambda(t)= -\lim_{N\rightarrow \infty} \frac{1}{N} \mathcal L(t) \nonumber
\\ \approx 
2    \tilde r_0^2 / b^2 - 2\, \min\left[ \mbox{Re}\zeta_+(t),\mbox{Re}\zeta_-(t)  \right]   \nonumber \\ 
= \min[\lambda_+(t),\lambda_-(t)], 
\end{gather}

where 
\bea\label{app:lambda_pm}
\lambda_+(t) &=& \frac{2 \tilde r_0^2 b^2}{\left[ b^4 +\tan^2(\pi^2 t)\right]},\\
\lambda_-(t) &=& \frac{2 \tilde r_0^2 b^2}{ \left[ b^4 +\cot^2(\pi^2 t)\right]},
\eea
which is finite in the thermodynamic limit. 


\begin{thebibliography}{93}
\expandafter\ifx\csname natexlab\endcsname\relax\def\natexlab#1{#1}\fi
\expandafter\ifx\csname bibnamefont\endcsname\relax
  \def\bibnamefont#1{#1}\fi
\expandafter\ifx\csname bibfnamefont\endcsname\relax
  \def\bibfnamefont#1{#1}\fi
\expandafter\ifx\csname citenamefont\endcsname\relax
  \def\citenamefont#1{#1}\fi
\expandafter\ifx\csname url\endcsname\relax
  \def\url#1{\texttt{#1}}\fi
\expandafter\ifx\csname urlprefix\endcsname\relax\def\urlprefix{URL }\fi
\providecommand{\bibinfo}[2]{#2}
\providecommand{\eprint}[2][]{\url{#2}}

\bibitem[{\citenamefont{{Sacha} and {Zakrzewski}}(2018)}]{Sacha2017rev}
\bibinfo{author}{\bibfnamefont{K.}~\bibnamefont{{Sacha}}} \bibnamefont{and}
  \bibinfo{author}{\bibfnamefont{J.}~\bibnamefont{{Zakrzewski}}},
  \bibinfo{journal}{Rep. Prog. Phys.} \textbf{\bibinfo{volume}{81}},
  \bibinfo{pages}{016401} (\bibinfo{year}{2018}),
  \urlprefix\url{https://doi.org/10.1088/1361-6633/aa8b38}.

\bibitem[{\citenamefont{Sachdev}(2011)}]{Sachdev2011}
\bibinfo{author}{\bibfnamefont{S.}~\bibnamefont{Sachdev}},
  \emph{\bibinfo{title}{{Quantum Phase Transitions}}}
  (\bibinfo{publisher}{{Cambridge University Press}},
  \bibinfo{address}{{Cambridge}}, \bibinfo{year}{2011}).

\bibitem[{\citenamefont{Dziarmaga}(2010)}]{Dziarmaga2010}
\bibinfo{author}{\bibfnamefont{J.}~\bibnamefont{Dziarmaga}},
  \bibinfo{journal}{Advances in Physics} \textbf{\bibinfo{volume}{59}},
  \bibinfo{pages}{1063} (\bibinfo{year}{2010}),
  \eprint{https://doi.org/10.1080/00018732.2010.514702},
  \urlprefix\url{https://doi.org/10.1080/00018732.2010.514702}.

\bibitem[{\citenamefont{Kibble}(1980)}]{Kibble1980}
\bibinfo{author}{\bibfnamefont{T.}~\bibnamefont{Kibble}},
  \bibinfo{journal}{Physics Reports} \textbf{\bibinfo{volume}{67}},
  \bibinfo{pages}{183 } (\bibinfo{year}{1980}), ISSN \bibinfo{issn}{0370-1573},
  \urlprefix\url{http://www.sciencedirect.com/science/article/pii/0370157380900915}.

\bibitem[{\citenamefont{Zurek}(1996)}]{Zurek1996}
\bibinfo{author}{\bibfnamefont{W.}~\bibnamefont{Zurek}},
  \bibinfo{journal}{Physics Reports} \textbf{\bibinfo{volume}{276}},
  \bibinfo{pages}{177 } (\bibinfo{year}{1996}), ISSN \bibinfo{issn}{0370-1573},
  \urlprefix\url{http://www.sciencedirect.com/science/article/pii/S0370157396000099}.

\bibitem[{\citenamefont{Damski}(2005)}]{Damski2005}
\bibinfo{author}{\bibfnamefont{B.}~\bibnamefont{Damski}},
  \bibinfo{journal}{Phys. Rev. Lett.} \textbf{\bibinfo{volume}{95}},
  \bibinfo{pages}{035701} (\bibinfo{year}{2005}),
  \urlprefix\url{https://link.aps.org/doi/10.1103/PhysRevLett.95.035701}.

\bibitem[{\citenamefont{Zurek et~al.}(2005)\citenamefont{Zurek, Dorner, and
  Zoller}}]{Zurek2005}
\bibinfo{author}{\bibfnamefont{W.~H.} \bibnamefont{Zurek}},
  \bibinfo{author}{\bibfnamefont{U.}~\bibnamefont{Dorner}}, \bibnamefont{and}
  \bibinfo{author}{\bibfnamefont{P.}~\bibnamefont{Zoller}},
  \bibinfo{journal}{Phys. Rev. Lett.} \textbf{\bibinfo{volume}{95}},
  \bibinfo{pages}{105701} (\bibinfo{year}{2005}),
  \urlprefix\url{https://link.aps.org/doi/10.1103/PhysRevLett.95.105701}.

\bibitem[{\citenamefont{Dziarmaga}(2005)}]{Dziarmaga2005}
\bibinfo{author}{\bibfnamefont{J.}~\bibnamefont{Dziarmaga}},
  \bibinfo{journal}{Phys. Rev. Lett.} \textbf{\bibinfo{volume}{95}},
  \bibinfo{pages}{245701} (\bibinfo{year}{2005}),
  \urlprefix\url{https://link.aps.org/doi/10.1103/PhysRevLett.95.245701}.

\bibitem[{\citenamefont{Polkovnikov}(2005)}]{Polkovnikov2005}
\bibinfo{author}{\bibfnamefont{A.}~\bibnamefont{Polkovnikov}},
  \bibinfo{journal}{Phys. Rev. B} \textbf{\bibinfo{volume}{72}},
  \bibinfo{pages}{161201} (\bibinfo{year}{2005}),
  \urlprefix\url{https://link.aps.org/doi/10.1103/PhysRevB.72.161201}.

\bibitem[{\citenamefont{Uhlmann et~al.}(2007)\citenamefont{Uhlmann,
  Sch\"utzhold, and Fischer}}]{Uhlmann2007}
\bibinfo{author}{\bibfnamefont{M.}~\bibnamefont{Uhlmann}},
  \bibinfo{author}{\bibfnamefont{R.}~\bibnamefont{Sch\"utzhold}},
  \bibnamefont{and} \bibinfo{author}{\bibfnamefont{U.~R.}
  \bibnamefont{Fischer}}, \bibinfo{journal}{Phys. Rev. Lett.}
  \textbf{\bibinfo{volume}{99}}, \bibinfo{pages}{120407}
  (\bibinfo{year}{2007}),
  \urlprefix\url{https://link.aps.org/doi/10.1103/PhysRevLett.99.120407}.

\bibitem[{\citenamefont{Uhlmann et~al.}(2010)\citenamefont{Uhlmann,
  Schützhold, and Fischer}}]{Uhlmann2010}
\bibinfo{author}{\bibfnamefont{M.}~\bibnamefont{Uhlmann}},
  \bibinfo{author}{\bibfnamefont{R.}~\bibnamefont{Schützhold}},
  \bibnamefont{and} \bibinfo{author}{\bibfnamefont{U.~R.}
  \bibnamefont{Fischer}}, \bibinfo{journal}{New Journal of Physics}
  \textbf{\bibinfo{volume}{12}}, \bibinfo{pages}{095020}
  (\bibinfo{year}{2010}),
  \urlprefix\url{http://stacks.iop.org/1367-2630/12/i=9/a=095020}.

\bibitem[{\citenamefont{\ifmmode~\acute{S}\else \'{S}\fi{}wis\l{}ocki
  et~al.}(2013)\citenamefont{\ifmmode~\acute{S}\else \'{S}\fi{}wis\l{}ocki,
  Witkowska, Dziarmaga, and Matuszewski}}]{Swislocki2013}
\bibinfo{author}{\bibfnamefont{T.}~\bibnamefont{\ifmmode~\acute{S}\else
  \'{S}\fi{}wis\l{}ocki}},
  \bibinfo{author}{\bibfnamefont{E.}~\bibnamefont{Witkowska}},
  \bibinfo{author}{\bibfnamefont{J.}~\bibnamefont{Dziarmaga}},
  \bibnamefont{and}
  \bibinfo{author}{\bibfnamefont{M.}~\bibnamefont{Matuszewski}},
  \bibinfo{journal}{Phys. Rev. Lett.} \textbf{\bibinfo{volume}{110}},
  \bibinfo{pages}{045303} (\bibinfo{year}{2013}),
  \urlprefix\url{https://link.aps.org/doi/10.1103/PhysRevLett.110.045303}.

\bibitem[{\citenamefont{Witkowska et~al.}(2013)\citenamefont{Witkowska,
  Dziarmaga, \ifmmode~\acute{S}\else \'{S}\fi{}wis\l{}ocki, and
  Matuszewski}}]{Witkowska2013}
\bibinfo{author}{\bibfnamefont{E.}~\bibnamefont{Witkowska}},
  \bibinfo{author}{\bibfnamefont{J.}~\bibnamefont{Dziarmaga}},
  \bibinfo{author}{\bibfnamefont{T.}~\bibnamefont{\ifmmode~\acute{S}\else
  \'{S}\fi{}wis\l{}ocki}}, \bibnamefont{and}
  \bibinfo{author}{\bibfnamefont{M.}~\bibnamefont{Matuszewski}},
  \bibinfo{journal}{Phys. Rev. B} \textbf{\bibinfo{volume}{88}},
  \bibinfo{pages}{054508} (\bibinfo{year}{2013}),
  \urlprefix\url{https://link.aps.org/doi/10.1103/PhysRevB.88.054508}.

\bibitem[{\citenamefont{Lacki and Damski}(2017)}]{Lacki2017}
\bibinfo{author}{\bibfnamefont{M.}~\bibnamefont{Lacki}} \bibnamefont{and}
  \bibinfo{author}{\bibfnamefont{B.}~\bibnamefont{Damski}},
  \bibinfo{journal}{Journal of Statistical Mechanics: Theory and Experiment}
  \textbf{\bibinfo{volume}{2017}}, \bibinfo{pages}{103105}
  (\bibinfo{year}{2017}),
  \urlprefix\url{http://stacks.iop.org/1742-5468/2017/i=10/a=103105}.

\bibitem[{\citenamefont{Bialonczyk and Damski}(2018)}]{Bialonczyk2018}
\bibinfo{author}{\bibfnamefont{M.}~\bibnamefont{Bialonczyk}} \bibnamefont{and}
  \bibinfo{author}{\bibfnamefont{B.}~\bibnamefont{Damski}},
  \bibinfo{journal}{ArXiv e-prints}  (\bibinfo{year}{2018}),
  \eprint{1803.05208}.

\bibitem[{\citenamefont{Heyl et~al.}(2013)\citenamefont{Heyl, Polkovnikov, and
  Kehrein}}]{Heyl2013}
\bibinfo{author}{\bibfnamefont{M.}~\bibnamefont{Heyl}},
  \bibinfo{author}{\bibfnamefont{A.}~\bibnamefont{Polkovnikov}},
  \bibnamefont{and} \bibinfo{author}{\bibfnamefont{S.}~\bibnamefont{Kehrein}},
  \bibinfo{journal}{Phys. Rev. Lett.} \textbf{\bibinfo{volume}{110}},
  \bibinfo{pages}{135704} (\bibinfo{year}{2013}),
  \urlprefix\url{https://link.aps.org/doi/10.1103/PhysRevLett.110.135704}.

\bibitem[{\citenamefont{Hickey et~al.}(2014)\citenamefont{Hickey, Genway, and
  Garrahan}}]{Hickey2014}
\bibinfo{author}{\bibfnamefont{J.~M.} \bibnamefont{Hickey}},
  \bibinfo{author}{\bibfnamefont{S.}~\bibnamefont{Genway}}, \bibnamefont{and}
  \bibinfo{author}{\bibfnamefont{J.~P.} \bibnamefont{Garrahan}},
  \bibinfo{journal}{Phys. Rev. B} \textbf{\bibinfo{volume}{89}},
  \bibinfo{pages}{054301} (\bibinfo{year}{2014}),
  \urlprefix\url{https://link.aps.org/doi/10.1103/PhysRevB.89.054301}.

\bibitem[{\citenamefont{Heyl}(2015)}]{Heyl2015}
\bibinfo{author}{\bibfnamefont{M.}~\bibnamefont{Heyl}}, \bibinfo{journal}{Phys.
  Rev. Lett.} \textbf{\bibinfo{volume}{115}}, \bibinfo{pages}{140602}
  (\bibinfo{year}{2015}),
  \urlprefix\url{https://link.aps.org/doi/10.1103/PhysRevLett.115.140602}.

\bibitem[{\citenamefont{Budich and Heyl}(2016)}]{Budich2016}
\bibinfo{author}{\bibfnamefont{J.~C.} \bibnamefont{Budich}} \bibnamefont{and}
  \bibinfo{author}{\bibfnamefont{M.}~\bibnamefont{Heyl}},
  \bibinfo{journal}{Phys. Rev. B} \textbf{\bibinfo{volume}{93}},
  \bibinfo{pages}{085416} (\bibinfo{year}{2016}),
  \urlprefix\url{https://link.aps.org/doi/10.1103/PhysRevB.93.085416}.

\bibitem[{\citenamefont{Sharma et~al.}(2016)\citenamefont{Sharma, Divakaran,
  Polkovnikov, and Dutta}}]{Sharma2016}
\bibinfo{author}{\bibfnamefont{S.}~\bibnamefont{Sharma}},
  \bibinfo{author}{\bibfnamefont{U.}~\bibnamefont{Divakaran}},
  \bibinfo{author}{\bibfnamefont{A.}~\bibnamefont{Polkovnikov}},
  \bibnamefont{and} \bibinfo{author}{\bibfnamefont{A.}~\bibnamefont{Dutta}},
  \bibinfo{journal}{Phys. Rev. B} \textbf{\bibinfo{volume}{93}},
  \bibinfo{pages}{144306} (\bibinfo{year}{2016}),
  \urlprefix\url{https://link.aps.org/doi/10.1103/PhysRevB.93.144306}.

\bibitem[{\citenamefont{\ifmmode \check{Z}\else
  \v{Z}\fi{}unkovi\ifmmode~\check{c}\else \v{c}\fi{}
  et~al.}(2018)\citenamefont{\ifmmode \check{Z}\else
  \v{Z}\fi{}unkovi\ifmmode~\check{c}\else \v{c}\fi{}, Heyl, Knap, and
  Silva}}]{Zunkovic2016}
\bibinfo{author}{\bibfnamefont{B.}~\bibnamefont{\ifmmode \check{Z}\else
  \v{Z}\fi{}unkovi\ifmmode~\check{c}\else \v{c}\fi{}}},
  \bibinfo{author}{\bibfnamefont{M.}~\bibnamefont{Heyl}},
  \bibinfo{author}{\bibfnamefont{M.}~\bibnamefont{Knap}}, \bibnamefont{and}
  \bibinfo{author}{\bibfnamefont{A.}~\bibnamefont{Silva}},
  \bibinfo{journal}{Phys. Rev. Lett.} \textbf{\bibinfo{volume}{120}},
  \bibinfo{pages}{130601} (\bibinfo{year}{2018}),
  \urlprefix\url{https://link.aps.org/doi/10.1103/PhysRevLett.120.130601}.

\bibitem[{\citenamefont{Bhattacharya and Dutta}(2017)}]{Bhattacharya2017a}
\bibinfo{author}{\bibfnamefont{U.}~\bibnamefont{Bhattacharya}}
  \bibnamefont{and} \bibinfo{author}{\bibfnamefont{A.}~\bibnamefont{Dutta}},
  \bibinfo{journal}{Phys. Rev. B} \textbf{\bibinfo{volume}{96}},
  \bibinfo{pages}{014302} (\bibinfo{year}{2017}),
  \urlprefix\url{https://link.aps.org/doi/10.1103/PhysRevB.96.014302}.

\bibitem[{\citenamefont{Bhattacharya et~al.}(2017)\citenamefont{Bhattacharya,
  Bandyopadhyay, and Dutta}}]{Bhattacharya2017b}
\bibinfo{author}{\bibfnamefont{U.}~\bibnamefont{Bhattacharya}},
  \bibinfo{author}{\bibfnamefont{S.}~\bibnamefont{Bandyopadhyay}},
  \bibnamefont{and} \bibinfo{author}{\bibfnamefont{A.}~\bibnamefont{Dutta}},
  \bibinfo{journal}{Phys. Rev. B} \textbf{\bibinfo{volume}{96}},
  \bibinfo{pages}{180303} (\bibinfo{year}{2017}),
  \urlprefix\url{https://link.aps.org/doi/10.1103/PhysRevB.96.180303}.

\bibitem[{\citenamefont{Karrasch and Schuricht}(2017)}]{Karrasch2017}
\bibinfo{author}{\bibfnamefont{C.}~\bibnamefont{Karrasch}} \bibnamefont{and}
  \bibinfo{author}{\bibfnamefont{D.}~\bibnamefont{Schuricht}},
  \bibinfo{journal}{Phys. Rev. B} \textbf{\bibinfo{volume}{95}},
  \bibinfo{pages}{075143} (\bibinfo{year}{2017}),
  \urlprefix\url{https://link.aps.org/doi/10.1103/PhysRevB.95.075143}.

\bibitem[{\citenamefont{Heyl and Budich}(2017)}]{Heyl2017}
\bibinfo{author}{\bibfnamefont{M.}~\bibnamefont{Heyl}} \bibnamefont{and}
  \bibinfo{author}{\bibfnamefont{J.~C.} \bibnamefont{Budich}},
  \bibinfo{journal}{Phys. Rev. B} \textbf{\bibinfo{volume}{96}},
  \bibinfo{pages}{180304} (\bibinfo{year}{2017}),
  \urlprefix\url{https://link.aps.org/doi/10.1103/PhysRevB.96.180304}.

\bibitem[{\citenamefont{Halimeh and Zauner-Stauber}(2017)}]{Halimeh2017}
\bibinfo{author}{\bibfnamefont{J.~C.} \bibnamefont{Halimeh}} \bibnamefont{and}
  \bibinfo{author}{\bibfnamefont{V.}~\bibnamefont{Zauner-Stauber}},
  \bibinfo{journal}{Phys. Rev. B} \textbf{\bibinfo{volume}{96}},
  \bibinfo{pages}{134427} (\bibinfo{year}{2017}),
  \urlprefix\url{https://link.aps.org/doi/10.1103/PhysRevB.96.134427}.

\bibitem[{\citenamefont{Zauner-Stauber and Halimeh}(2017)}]{Zauner-Stauber2017}
\bibinfo{author}{\bibfnamefont{V.}~\bibnamefont{Zauner-Stauber}}
  \bibnamefont{and} \bibinfo{author}{\bibfnamefont{J.~C.}
  \bibnamefont{Halimeh}}, \bibinfo{journal}{Phys. Rev. E}
  \textbf{\bibinfo{volume}{96}}, \bibinfo{pages}{062118}
  (\bibinfo{year}{2017}),
  \urlprefix\url{https://link.aps.org/doi/10.1103/PhysRevE.96.062118}.

\bibitem[{\citenamefont{Homrighausen et~al.}(2017)\citenamefont{Homrighausen,
  Abeling, Zauner-Stauber, and Halimeh}}]{Homrighausen2017}
\bibinfo{author}{\bibfnamefont{I.}~\bibnamefont{Homrighausen}},
  \bibinfo{author}{\bibfnamefont{N.~O.} \bibnamefont{Abeling}},
  \bibinfo{author}{\bibfnamefont{V.}~\bibnamefont{Zauner-Stauber}},
  \bibnamefont{and} \bibinfo{author}{\bibfnamefont{J.~C.}
  \bibnamefont{Halimeh}}, \bibinfo{journal}{Phys. Rev. B}
  \textbf{\bibinfo{volume}{96}}, \bibinfo{pages}{104436}
  (\bibinfo{year}{2017}),
  \urlprefix\url{https://link.aps.org/doi/10.1103/PhysRevB.96.104436}.

\bibitem[{\citenamefont{Lang et~al.}(2018)\citenamefont{Lang, Frank, and
  Halimeh}}]{Lang2017}
\bibinfo{author}{\bibfnamefont{J.}~\bibnamefont{Lang}},
  \bibinfo{author}{\bibfnamefont{B.}~\bibnamefont{Frank}}, \bibnamefont{and}
  \bibinfo{author}{\bibfnamefont{J.~C.} \bibnamefont{Halimeh}},
  \bibinfo{journal}{Phys. Rev. B} \textbf{\bibinfo{volume}{97}},
  \bibinfo{pages}{174401} (\bibinfo{year}{2018}),
  \urlprefix\url{https://link.aps.org/doi/10.1103/PhysRevB.97.174401}.

\bibitem[{\citenamefont{Weidinger et~al.}(2017)\citenamefont{Weidinger, Heyl,
  Silva, and Knap}}]{Weidinger2017}
\bibinfo{author}{\bibfnamefont{S.~A.} \bibnamefont{Weidinger}},
  \bibinfo{author}{\bibfnamefont{M.}~\bibnamefont{Heyl}},
  \bibinfo{author}{\bibfnamefont{A.}~\bibnamefont{Silva}}, \bibnamefont{and}
  \bibinfo{author}{\bibfnamefont{M.}~\bibnamefont{Knap}},
  \bibinfo{journal}{Phys. Rev. B} \textbf{\bibinfo{volume}{96}},
  \bibinfo{pages}{134313} (\bibinfo{year}{2017}),
  \urlprefix\url{https://link.aps.org/doi/10.1103/PhysRevB.96.134313}.

\bibitem[{\citenamefont{Chichinadze and Rubtsov}(2017)}]{Chichinadze2017}
\bibinfo{author}{\bibfnamefont{D.~V.} \bibnamefont{Chichinadze}}
  \bibnamefont{and} \bibinfo{author}{\bibfnamefont{A.~N.}
  \bibnamefont{Rubtsov}}, \bibinfo{journal}{Phys. Rev. B}
  \textbf{\bibinfo{volume}{95}}, \bibinfo{pages}{180302}
  (\bibinfo{year}{2017}),
  \urlprefix\url{https://link.aps.org/doi/10.1103/PhysRevB.95.180302}.

\bibitem[{\citenamefont{Lerose et~al.}(2018)\citenamefont{Lerose, Marino,
  \ifmmode \check{Z}\else \v{Z}\fi{}unkovi\ifmmode~\check{c}\else \v{c}\fi{},
  Gambassi, and Silva}}]{Lerose2018}
\bibinfo{author}{\bibfnamefont{A.}~\bibnamefont{Lerose}},
  \bibinfo{author}{\bibfnamefont{J.}~\bibnamefont{Marino}},
  \bibinfo{author}{\bibfnamefont{B.}~\bibnamefont{\ifmmode \check{Z}\else
  \v{Z}\fi{}unkovi\ifmmode~\check{c}\else \v{c}\fi{}}},
  \bibinfo{author}{\bibfnamefont{A.}~\bibnamefont{Gambassi}}, \bibnamefont{and}
  \bibinfo{author}{\bibfnamefont{A.}~\bibnamefont{Silva}},
  \bibinfo{journal}{Phys. Rev. Lett.} \textbf{\bibinfo{volume}{120}},
  \bibinfo{pages}{130603} (\bibinfo{year}{2018}),
  \urlprefix\url{https://link.aps.org/doi/10.1103/PhysRevLett.120.130603}.

\bibitem[{\citenamefont{Mera et~al.}(2018)\citenamefont{Mera, Vlachou,
  Paunkovi\ifmmode~\acute{c}\else \'{c}\fi{}, Vieira, and Viyuela}}]{Mera2018}
\bibinfo{author}{\bibfnamefont{B.}~\bibnamefont{Mera}},
  \bibinfo{author}{\bibfnamefont{C.}~\bibnamefont{Vlachou}},
  \bibinfo{author}{\bibfnamefont{N.}~\bibnamefont{Paunkovi\ifmmode~\acute{c}\else
  \'{c}\fi{}}}, \bibinfo{author}{\bibfnamefont{V.~R.} \bibnamefont{Vieira}},
  \bibnamefont{and} \bibinfo{author}{\bibfnamefont{O.}~\bibnamefont{Viyuela}},
  \bibinfo{journal}{Phys. Rev. B} \textbf{\bibinfo{volume}{97}},
  \bibinfo{pages}{094110} (\bibinfo{year}{2018}),
  \urlprefix\url{https://link.aps.org/doi/10.1103/PhysRevB.97.094110}.

\bibitem[{\citenamefont{Piroli et~al.}(2018)\citenamefont{Piroli, Pozsgay, and
  Vernier}}]{Piroli2018}
\bibinfo{author}{\bibfnamefont{L.}~\bibnamefont{Piroli}},
  \bibinfo{author}{\bibfnamefont{B.}~\bibnamefont{Pozsgay}}, \bibnamefont{and}
  \bibinfo{author}{\bibfnamefont{E.}~\bibnamefont{Vernier}},
  \bibinfo{journal}{Nuclear Physics B} \textbf{\bibinfo{volume}{933}},
  \bibinfo{pages}{454 } (\bibinfo{year}{2018}), ISSN \bibinfo{issn}{0550-3213},
  \urlprefix\url{http://www.sciencedirect.com/science/article/pii/S055032131830172X}.

\bibitem[{\citenamefont{Andraschko and Sirker}(2014)}]{Andraschko2014}
\bibinfo{author}{\bibfnamefont{F.}~\bibnamefont{Andraschko}} \bibnamefont{and}
  \bibinfo{author}{\bibfnamefont{J.}~\bibnamefont{Sirker}},
  \bibinfo{journal}{Phys. Rev. B} \textbf{\bibinfo{volume}{89}},
  \bibinfo{pages}{125120} (\bibinfo{year}{2014}),
  \urlprefix\url{https://link.aps.org/doi/10.1103/PhysRevB.89.125120}.

\bibitem[{\citenamefont{Vajna and D\'ora}(2014)}]{Vajna2014}
\bibinfo{author}{\bibfnamefont{S.}~\bibnamefont{Vajna}} \bibnamefont{and}
  \bibinfo{author}{\bibfnamefont{B.}~\bibnamefont{D\'ora}},
  \bibinfo{journal}{Phys. Rev. B} \textbf{\bibinfo{volume}{89}},
  \bibinfo{pages}{161105} (\bibinfo{year}{2014}).

\bibitem[{\citenamefont{Jurcevic et~al.}(2017)\citenamefont{Jurcevic, Shen,
  Hauke, Maier, Brydges, Hempel, Lanyon, Heyl, Blatt, and Roos}}]{Jurcevic2017}
\bibinfo{author}{\bibfnamefont{P.}~\bibnamefont{Jurcevic}},
  \bibinfo{author}{\bibfnamefont{H.}~\bibnamefont{Shen}},
  \bibinfo{author}{\bibfnamefont{P.}~\bibnamefont{Hauke}},
  \bibinfo{author}{\bibfnamefont{C.}~\bibnamefont{Maier}},
  \bibinfo{author}{\bibfnamefont{T.}~\bibnamefont{Brydges}},
  \bibinfo{author}{\bibfnamefont{C.}~\bibnamefont{Hempel}},
  \bibinfo{author}{\bibfnamefont{B.~P.} \bibnamefont{Lanyon}},
  \bibinfo{author}{\bibfnamefont{M.}~\bibnamefont{Heyl}},
  \bibinfo{author}{\bibfnamefont{R.}~\bibnamefont{Blatt}}, \bibnamefont{and}
  \bibinfo{author}{\bibfnamefont{C.~F.} \bibnamefont{Roos}},
  \bibinfo{journal}{Phys. Rev. Lett.} \textbf{\bibinfo{volume}{119}},
  \bibinfo{pages}{080501} (\bibinfo{year}{2017}),
  \urlprefix\url{https://link.aps.org/doi/10.1103/PhysRevLett.119.080501}.

\bibitem[{\citenamefont{Fl\"aschner et~al.}(2018)\citenamefont{Fl\"aschner,
  Vogel, Tarnowski, Rem, L\"uhmann, Heyl, Budich, Mathey, Sengstock, and
  Weitenberg}}]{Flaschner2018}
\bibinfo{author}{\bibfnamefont{N.}~\bibnamefont{Fl\"aschner}},
  \bibinfo{author}{\bibfnamefont{D.}~\bibnamefont{Vogel}},
  \bibinfo{author}{\bibfnamefont{M.}~\bibnamefont{Tarnowski}},
  \bibinfo{author}{\bibfnamefont{B.~S.} \bibnamefont{Rem}},
  \bibinfo{author}{\bibfnamefont{D.-S.} \bibnamefont{L\"uhmann}},
  \bibinfo{author}{\bibfnamefont{M.}~\bibnamefont{Heyl}},
  \bibinfo{author}{\bibfnamefont{J.~C.} \bibnamefont{Budich}},
  \bibinfo{author}{\bibfnamefont{L.}~\bibnamefont{Mathey}},
  \bibinfo{author}{\bibfnamefont{K.}~\bibnamefont{Sengstock}},
  \bibnamefont{and}
  \bibinfo{author}{\bibfnamefont{C.}~\bibnamefont{Weitenberg}},
  \bibinfo{journal}{Nature Physics} \textbf{\bibinfo{volume}{14}},
  \bibinfo{pages}{265} (\bibinfo{year}{2018}).

\bibitem[{\citenamefont{Heyl}(2018)}]{Heyl2018_review}
\bibinfo{author}{\bibfnamefont{M.}~\bibnamefont{Heyl}},
  \bibinfo{journal}{Reports on Progress in Physics}
  \textbf{\bibinfo{volume}{81}}, \bibinfo{pages}{054001}
  (\bibinfo{year}{2018}).

\bibitem[{\citenamefont{Wilczek}(2012)}]{Wilczek2012}
\bibinfo{author}{\bibfnamefont{F.}~\bibnamefont{Wilczek}},
  \bibinfo{journal}{Phys. Rev. Lett.} \textbf{\bibinfo{volume}{109}},
  \bibinfo{pages}{160401} (\bibinfo{year}{2012}),
  \urlprefix\url{http://link.aps.org/doi/10.1103/PhysRevLett.109.160401}.

\bibitem[{\citenamefont{Bruno}(2013)}]{Bruno2013b}
\bibinfo{author}{\bibfnamefont{P.}~\bibnamefont{Bruno}},
  \bibinfo{journal}{Phys. Rev. Lett.} \textbf{\bibinfo{volume}{111}},
  \bibinfo{pages}{070402} (\bibinfo{year}{2013}),
  \urlprefix\url{http://link.aps.org/doi/10.1103/PhysRevLett.111.070402}.

\bibitem[{\citenamefont{Watanabe and Oshikawa}(2015)}]{Watanabe2015}
\bibinfo{author}{\bibfnamefont{H.}~\bibnamefont{Watanabe}} \bibnamefont{and}
  \bibinfo{author}{\bibfnamefont{M.}~\bibnamefont{Oshikawa}},
  \bibinfo{journal}{Phys. Rev. Lett.} \textbf{\bibinfo{volume}{114}},
  \bibinfo{pages}{251603} (\bibinfo{year}{2015}),
  \urlprefix\url{http://link.aps.org/doi/10.1103/PhysRevLett.114.251603}.

\bibitem[{\citenamefont{Syrwid et~al.}(2017)\citenamefont{Syrwid, Zakrzewski,
  and Sacha}}]{Syrwid2017}
\bibinfo{author}{\bibfnamefont{A.}~\bibnamefont{Syrwid}},
  \bibinfo{author}{\bibfnamefont{J.}~\bibnamefont{Zakrzewski}},
  \bibnamefont{and} \bibinfo{author}{\bibfnamefont{K.}~\bibnamefont{Sacha}},
  \bibinfo{journal}{Phys. Rev. Lett.} \textbf{\bibinfo{volume}{119}},
  \bibinfo{pages}{250602} (\bibinfo{year}{2017}),
  \urlprefix\url{https://link.aps.org/doi/10.1103/PhysRevLett.119.250602}.

\bibitem[{\citenamefont{Iemini et~al.}(2018)\citenamefont{Iemini, Russomanno,
  Keeling, Schir\`o, Dalmonte, and Fazio}}]{Iemini2017}
\bibinfo{author}{\bibfnamefont{F.}~\bibnamefont{Iemini}},
  \bibinfo{author}{\bibfnamefont{A.}~\bibnamefont{Russomanno}},
  \bibinfo{author}{\bibfnamefont{J.}~\bibnamefont{Keeling}},
  \bibinfo{author}{\bibfnamefont{M.}~\bibnamefont{Schir\`o}},
  \bibinfo{author}{\bibfnamefont{M.}~\bibnamefont{Dalmonte}}, \bibnamefont{and}
  \bibinfo{author}{\bibfnamefont{R.}~\bibnamefont{Fazio}},
  \bibinfo{journal}{Phys. Rev. Lett.} \textbf{\bibinfo{volume}{121}},
  \bibinfo{pages}{035301} (\bibinfo{year}{2018}),
  \urlprefix\url{https://link.aps.org/doi/10.1103/PhysRevLett.121.035301}.

\bibitem[{\citenamefont{Huang et~al.}(2018{\natexlab{a}})\citenamefont{Huang,
  Li, and Yin}}]{Huang2017a}
\bibinfo{author}{\bibfnamefont{Y.}~\bibnamefont{Huang}},
  \bibinfo{author}{\bibfnamefont{T.}~\bibnamefont{Li}}, \bibnamefont{and}
  \bibinfo{author}{\bibfnamefont{Z.-q.} \bibnamefont{Yin}},
  \bibinfo{journal}{Phys. Rev. A} \textbf{\bibinfo{volume}{97}},
  \bibinfo{pages}{012115} (\bibinfo{year}{2018}{\natexlab{a}}),
  \urlprefix\url{https://link.aps.org/doi/10.1103/PhysRevA.97.012115}.

\bibitem[{\citenamefont{{Prokof'ev} and {Svistunov}}(2017)}]{Prokofev2017}
\bibinfo{author}{\bibfnamefont{N.~V.} \bibnamefont{{Prokof'ev}}}
  \bibnamefont{and} \bibinfo{author}{\bibfnamefont{B.~V.}
  \bibnamefont{{Svistunov}}}, \bibinfo{journal}{ArXiv e-prints}
  (\bibinfo{year}{2017}), \eprint{1710.00721}.

\bibitem[{\citenamefont{Sacha}(2015{\natexlab{a}})}]{Sacha2015}
\bibinfo{author}{\bibfnamefont{K.}~\bibnamefont{Sacha}},
  \bibinfo{journal}{Phys. Rev. A} \textbf{\bibinfo{volume}{91}},
  \bibinfo{pages}{033617} (\bibinfo{year}{2015}{\natexlab{a}}),
  \urlprefix\url{http://link.aps.org/doi/10.1103/PhysRevA.91.033617}.

\bibitem[{\citenamefont{Khemani et~al.}(2016)\citenamefont{Khemani, Lazarides,
  Moessner, and Sondhi}}]{Khemani16}
\bibinfo{author}{\bibfnamefont{V.}~\bibnamefont{Khemani}},
  \bibinfo{author}{\bibfnamefont{A.}~\bibnamefont{Lazarides}},
  \bibinfo{author}{\bibfnamefont{R.}~\bibnamefont{Moessner}}, \bibnamefont{and}
  \bibinfo{author}{\bibfnamefont{S.~L.} \bibnamefont{Sondhi}},
  \bibinfo{journal}{Phys. Rev. Lett.} \textbf{\bibinfo{volume}{116}},
  \bibinfo{pages}{250401} (\bibinfo{year}{2016}),
  \urlprefix\url{http://link.aps.org/doi/10.1103/PhysRevLett.116.250401}.

\bibitem[{\citenamefont{Else et~al.}(2016)\citenamefont{Else, Bauer, and
  Nayak}}]{ElseFTC}
\bibinfo{author}{\bibfnamefont{D.~V.} \bibnamefont{Else}},
  \bibinfo{author}{\bibfnamefont{B.}~\bibnamefont{Bauer}}, \bibnamefont{and}
  \bibinfo{author}{\bibfnamefont{C.}~\bibnamefont{Nayak}},
  \bibinfo{journal}{Phys. Rev. Lett.} \textbf{\bibinfo{volume}{117}},
  \bibinfo{pages}{090402} (\bibinfo{year}{2016}),
  \urlprefix\url{http://link.aps.org/doi/10.1103/PhysRevLett.117.090402}.

\bibitem[{\citenamefont{Yao et~al.}(2017)\citenamefont{Yao, Potter, Potirniche,
  and Vishwanath}}]{Yao2017}
\bibinfo{author}{\bibfnamefont{N.~Y.} \bibnamefont{Yao}},
  \bibinfo{author}{\bibfnamefont{A.~C.} \bibnamefont{Potter}},
  \bibinfo{author}{\bibfnamefont{I.-D.} \bibnamefont{Potirniche}},
  \bibnamefont{and}
  \bibinfo{author}{\bibfnamefont{A.}~\bibnamefont{Vishwanath}},
  \bibinfo{journal}{Phys. Rev. Lett.} \textbf{\bibinfo{volume}{118}},
  \bibinfo{pages}{030401} (\bibinfo{year}{2017}),
  \urlprefix\url{http://link.aps.org/doi/10.1103/PhysRevLett.118.030401}.

\bibitem[{\citenamefont{Lazarides and Moessner}(2017)}]{Lazarides2017}
\bibinfo{author}{\bibfnamefont{A.}~\bibnamefont{Lazarides}} \bibnamefont{and}
  \bibinfo{author}{\bibfnamefont{R.}~\bibnamefont{Moessner}},
  \bibinfo{journal}{Phys. Rev. B} \textbf{\bibinfo{volume}{95}},
  \bibinfo{pages}{195135} (\bibinfo{year}{2017}),
  \urlprefix\url{https://link.aps.org/doi/10.1103/PhysRevB.95.195135}.

\bibitem[{\citenamefont{Russomanno et~al.}(2017)\citenamefont{Russomanno,
  Iemini, Dalmonte, and Fazio}}]{Russomanno2017}
\bibinfo{author}{\bibfnamefont{A.}~\bibnamefont{Russomanno}},
  \bibinfo{author}{\bibfnamefont{F.}~\bibnamefont{Iemini}},
  \bibinfo{author}{\bibfnamefont{M.}~\bibnamefont{Dalmonte}}, \bibnamefont{and}
  \bibinfo{author}{\bibfnamefont{R.}~\bibnamefont{Fazio}},
  \bibinfo{journal}{Phys. Rev. B} \textbf{\bibinfo{volume}{95}},
  \bibinfo{pages}{214307} (\bibinfo{year}{2017}),
  \urlprefix\url{https://link.aps.org/doi/10.1103/PhysRevB.95.214307}.

\bibitem[{\citenamefont{Zeng and Sheng}(2017)}]{Zeng2017}
\bibinfo{author}{\bibfnamefont{T.-S.} \bibnamefont{Zeng}} \bibnamefont{and}
  \bibinfo{author}{\bibfnamefont{D.~N.} \bibnamefont{Sheng}},
  \bibinfo{journal}{Phys. Rev. B} \textbf{\bibinfo{volume}{96}},
  \bibinfo{pages}{094202} (\bibinfo{year}{2017}),
  \urlprefix\url{https://link.aps.org/doi/10.1103/PhysRevB.96.094202}.

\bibitem[{\citenamefont{Nakatsugawa et~al.}(2017)\citenamefont{Nakatsugawa,
  Fujii, and Tanda}}]{Nakatsugawa2017}
\bibinfo{author}{\bibfnamefont{K.}~\bibnamefont{Nakatsugawa}},
  \bibinfo{author}{\bibfnamefont{T.}~\bibnamefont{Fujii}}, \bibnamefont{and}
  \bibinfo{author}{\bibfnamefont{S.}~\bibnamefont{Tanda}},
  \bibinfo{journal}{Phys. Rev. B} \textbf{\bibinfo{volume}{96}},
  \bibinfo{pages}{094308} (\bibinfo{year}{2017}),
  \urlprefix\url{https://link.aps.org/doi/10.1103/PhysRevB.96.094308}.

\bibitem[{\citenamefont{Ho et~al.}(2017)\citenamefont{Ho, Choi, Lukin, and
  Abanin}}]{Ho2017}
\bibinfo{author}{\bibfnamefont{W.~W.} \bibnamefont{Ho}},
  \bibinfo{author}{\bibfnamefont{S.}~\bibnamefont{Choi}},
  \bibinfo{author}{\bibfnamefont{M.~D.} \bibnamefont{Lukin}}, \bibnamefont{and}
  \bibinfo{author}{\bibfnamefont{D.~A.} \bibnamefont{Abanin}},
  \bibinfo{journal}{Phys. Rev. Lett.} \textbf{\bibinfo{volume}{119}},
  \bibinfo{pages}{010602} (\bibinfo{year}{2017}),
  \urlprefix\url{https://link.aps.org/doi/10.1103/PhysRevLett.119.010602}.

\bibitem[{\citenamefont{Huang et~al.}(2018{\natexlab{b}})\citenamefont{Huang,
  Wu, and Liu}}]{Huang2017}
\bibinfo{author}{\bibfnamefont{B.}~\bibnamefont{Huang}},
  \bibinfo{author}{\bibfnamefont{Y.-H.} \bibnamefont{Wu}}, \bibnamefont{and}
  \bibinfo{author}{\bibfnamefont{W.~V.} \bibnamefont{Liu}},
  \bibinfo{journal}{Phys. Rev. Lett.} \textbf{\bibinfo{volume}{120}},
  \bibinfo{pages}{110603} (\bibinfo{year}{2018}{\natexlab{b}}),
  \urlprefix\url{https://link.aps.org/doi/10.1103/PhysRevLett.120.110603}.

\bibitem[{\citenamefont{Gong et~al.}(2018)\citenamefont{Gong, Hamazaki, and
  Ueda}}]{Gong2017}
\bibinfo{author}{\bibfnamefont{Z.}~\bibnamefont{Gong}},
  \bibinfo{author}{\bibfnamefont{R.}~\bibnamefont{Hamazaki}}, \bibnamefont{and}
  \bibinfo{author}{\bibfnamefont{M.}~\bibnamefont{Ueda}},
  \bibinfo{journal}{Phys. Rev. Lett.} \textbf{\bibinfo{volume}{120}},
  \bibinfo{pages}{040404} (\bibinfo{year}{2018}),
  \urlprefix\url{https://link.aps.org/doi/10.1103/PhysRevLett.120.040404}.

\bibitem[{\citenamefont{Wang et~al.}(2018)\citenamefont{Wang, Xing, Carlo, and
  Poletti}}]{Wang2017}
\bibinfo{author}{\bibfnamefont{R.~R.~W.} \bibnamefont{Wang}},
  \bibinfo{author}{\bibfnamefont{B.}~\bibnamefont{Xing}},
  \bibinfo{author}{\bibfnamefont{G.~G.} \bibnamefont{Carlo}}, \bibnamefont{and}
  \bibinfo{author}{\bibfnamefont{D.}~\bibnamefont{Poletti}},
  \bibinfo{journal}{Phys. Rev. E} \textbf{\bibinfo{volume}{97}},
  \bibinfo{pages}{020202} (\bibinfo{year}{2018}),
  \urlprefix\url{https://link.aps.org/doi/10.1103/PhysRevE.97.020202}.

\bibitem[{\citenamefont{Tucker et~al.}(2018)\citenamefont{Tucker, Zhu,
  Lewis-Swan, Marino, Jimenez, Restrepo, and Rey}}]{Tucker2018}
\bibinfo{author}{\bibfnamefont{K.}~\bibnamefont{Tucker}},
  \bibinfo{author}{\bibfnamefont{B.}~\bibnamefont{Zhu}},
  \bibinfo{author}{\bibfnamefont{R.}~\bibnamefont{Lewis-Swan}},
  \bibinfo{author}{\bibfnamefont{J.}~\bibnamefont{Marino}},
  \bibinfo{author}{\bibfnamefont{F.}~\bibnamefont{Jimenez}},
  \bibinfo{author}{\bibfnamefont{J.}~\bibnamefont{Restrepo}}, \bibnamefont{and}
  \bibinfo{author}{\bibfnamefont{A.~M.} \bibnamefont{Rey}},
  \bibinfo{journal}{arXiv preprint arXiv:1805.03343}  (\bibinfo{year}{2018}).

\bibitem[{\citenamefont{Zhang et~al.}(2017)\citenamefont{Zhang, Hess,
  Kyprianidis, Becker, Lee, Smith, Pagano, Potirniche, Potter, Vishwanath
  et~al.}}]{Zhang2017}
\bibinfo{author}{\bibfnamefont{J.}~\bibnamefont{Zhang}},
  \bibinfo{author}{\bibfnamefont{P.~W.} \bibnamefont{Hess}},
  \bibinfo{author}{\bibfnamefont{A.}~\bibnamefont{Kyprianidis}},
  \bibinfo{author}{\bibfnamefont{P.}~\bibnamefont{Becker}},
  \bibinfo{author}{\bibfnamefont{A.}~\bibnamefont{Lee}},
  \bibinfo{author}{\bibfnamefont{J.}~\bibnamefont{Smith}},
  \bibinfo{author}{\bibfnamefont{G.}~\bibnamefont{Pagano}},
  \bibinfo{author}{\bibfnamefont{I.-D.} \bibnamefont{Potirniche}},
  \bibinfo{author}{\bibfnamefont{A.~C.} \bibnamefont{Potter}},
  \bibinfo{author}{\bibfnamefont{A.}~\bibnamefont{Vishwanath}},
  \bibnamefont{et~al.}, \bibinfo{journal}{Nature}
  \textbf{\bibinfo{volume}{543}}, \bibinfo{pages}{217} (\bibinfo{year}{2017}),
  ISSN \bibinfo{issn}{0028-0836}, \bibinfo{note}{letter},
  \urlprefix\url{http://dx.doi.org/10.1038/nature21413}.

\bibitem[{\citenamefont{Choi et~al.}(2017)\citenamefont{Choi, Choi, Landig,
  Kucsko, Zhou, Isoya, Jelezko, Onoda, Sumiya, Khemani et~al.}}]{Choi2017}
\bibinfo{author}{\bibfnamefont{S.}~\bibnamefont{Choi}},
  \bibinfo{author}{\bibfnamefont{J.}~\bibnamefont{Choi}},
  \bibinfo{author}{\bibfnamefont{R.}~\bibnamefont{Landig}},
  \bibinfo{author}{\bibfnamefont{G.}~\bibnamefont{Kucsko}},
  \bibinfo{author}{\bibfnamefont{H.}~\bibnamefont{Zhou}},
  \bibinfo{author}{\bibfnamefont{J.}~\bibnamefont{Isoya}},
  \bibinfo{author}{\bibfnamefont{F.}~\bibnamefont{Jelezko}},
  \bibinfo{author}{\bibfnamefont{S.}~\bibnamefont{Onoda}},
  \bibinfo{author}{\bibfnamefont{H.}~\bibnamefont{Sumiya}},
  \bibinfo{author}{\bibfnamefont{V.}~\bibnamefont{Khemani}},
  \bibnamefont{et~al.}, \bibinfo{journal}{Nature}
  \textbf{\bibinfo{volume}{543}}, \bibinfo{pages}{221} (\bibinfo{year}{2017}),
  ISSN \bibinfo{issn}{0028-0836}, \bibinfo{note}{letter},
  \urlprefix\url{http://dx.doi.org/10.1038/nature21426}.

\bibitem[{\citenamefont{Nayak}(2017)}]{Nayak2017}
\bibinfo{author}{\bibfnamefont{C.}~\bibnamefont{Nayak}},
  \bibinfo{journal}{Nature} \textbf{\bibinfo{volume}{543}},
  \bibinfo{pages}{185} (\bibinfo{year}{2017}), ISSN \bibinfo{issn}{0028-0836},
  \bibinfo{note}{news \& Views},
  \urlprefix\url{http://dx.doi.org/10.1038/543185a}.

\bibitem[{\citenamefont{Pal et~al.}(2018)\citenamefont{Pal, Nishad, Mahesh, and
  Sreejith}}]{Pal2018}
\bibinfo{author}{\bibfnamefont{S.}~\bibnamefont{Pal}},
  \bibinfo{author}{\bibfnamefont{N.}~\bibnamefont{Nishad}},
  \bibinfo{author}{\bibfnamefont{T.~S.} \bibnamefont{Mahesh}},
  \bibnamefont{and} \bibinfo{author}{\bibfnamefont{G.~J.}
  \bibnamefont{Sreejith}}, \bibinfo{journal}{Phys. Rev. Lett.}
  \textbf{\bibinfo{volume}{120}}, \bibinfo{pages}{180602}
  (\bibinfo{year}{2018}),
  \urlprefix\url{https://link.aps.org/doi/10.1103/PhysRevLett.120.180602}.

\bibitem[{\citenamefont{Rovny et~al.}(2018{\natexlab{a}})\citenamefont{Rovny,
  Blum, and Barrett}}]{Rovny2018}
\bibinfo{author}{\bibfnamefont{J.}~\bibnamefont{Rovny}},
  \bibinfo{author}{\bibfnamefont{R.~L.} \bibnamefont{Blum}}, \bibnamefont{and}
  \bibinfo{author}{\bibfnamefont{S.~E.} \bibnamefont{Barrett}},
  \bibinfo{journal}{Phys. Rev. Lett.} \textbf{\bibinfo{volume}{120}},
  \bibinfo{pages}{180603} (\bibinfo{year}{2018}{\natexlab{a}}),
  \urlprefix\url{https://link.aps.org/doi/10.1103/PhysRevLett.120.180603}.

\bibitem[{\citenamefont{Rovny et~al.}(2018{\natexlab{b}})\citenamefont{Rovny,
  Blum, and Barrett}}]{Rovny2018a}
\bibinfo{author}{\bibfnamefont{J.}~\bibnamefont{Rovny}},
  \bibinfo{author}{\bibfnamefont{R.~L.} \bibnamefont{Blum}}, \bibnamefont{and}
  \bibinfo{author}{\bibfnamefont{S.~E.} \bibnamefont{Barrett}},
  \bibinfo{journal}{Phys. Rev. B} \textbf{\bibinfo{volume}{97}},
  \bibinfo{pages}{184301} (\bibinfo{year}{2018}{\natexlab{b}}),
  \urlprefix\url{https://link.aps.org/doi/10.1103/PhysRevB.97.184301}.

\bibitem[{\citenamefont{Autti et~al.}(2018)\citenamefont{Autti, Eltsov, and
  Volovik}}]{Autti2018}
\bibinfo{author}{\bibfnamefont{S.}~\bibnamefont{Autti}},
  \bibinfo{author}{\bibfnamefont{V.~B.} \bibnamefont{Eltsov}},
  \bibnamefont{and} \bibinfo{author}{\bibfnamefont{G.~E.}
  \bibnamefont{Volovik}}, \bibinfo{journal}{Phys. Rev. Lett.}
  \textbf{\bibinfo{volume}{120}}, \bibinfo{pages}{215301}
  (\bibinfo{year}{2018}),
  \urlprefix\url{https://link.aps.org/doi/10.1103/PhysRevLett.120.215301}.

\bibitem[{\citenamefont{Giergiel
  et~al.}(2018{\natexlab{a}})\citenamefont{Giergiel, Kosior, Hannaford, and
  Sacha}}]{Giergiel2018}
\bibinfo{author}{\bibfnamefont{K.}~\bibnamefont{Giergiel}},
  \bibinfo{author}{\bibfnamefont{A.}~\bibnamefont{Kosior}},
  \bibinfo{author}{\bibfnamefont{P.}~\bibnamefont{Hannaford}},
  \bibnamefont{and} \bibinfo{author}{\bibfnamefont{K.}~\bibnamefont{Sacha}},
  \bibinfo{journal}{Phys. Rev. A} \textbf{\bibinfo{volume}{98}},
  \bibinfo{pages}{013613} (\bibinfo{year}{2018}{\natexlab{a}}),
  \urlprefix\url{https://link.aps.org/doi/10.1103/PhysRevA.98.013613}.

\bibitem[{\citenamefont{Kim et~al.}(2006)\citenamefont{Kim, Heo, Lee, Jang,
  Noh, Kim, and Jhe}}]{Kim2006}
\bibinfo{author}{\bibfnamefont{K.}~\bibnamefont{Kim}},
  \bibinfo{author}{\bibfnamefont{M.-S.} \bibnamefont{Heo}},
  \bibinfo{author}{\bibfnamefont{K.-H.} \bibnamefont{Lee}},
  \bibinfo{author}{\bibfnamefont{K.}~\bibnamefont{Jang}},
  \bibinfo{author}{\bibfnamefont{H.-R.} \bibnamefont{Noh}},
  \bibinfo{author}{\bibfnamefont{D.}~\bibnamefont{Kim}}, \bibnamefont{and}
  \bibinfo{author}{\bibfnamefont{W.}~\bibnamefont{Jhe}},
  \bibinfo{journal}{Phys. Rev. Lett.} \textbf{\bibinfo{volume}{96}},
  \bibinfo{pages}{150601} (\bibinfo{year}{2006}),
  \urlprefix\url{https://link.aps.org/doi/10.1103/PhysRevLett.96.150601}.

\bibitem[{\citenamefont{Heo et~al.}(2010)\citenamefont{Heo, Kim, Kim, Moon,
  Lee, Noh, Dykman, and Jhe}}]{Heo2010}
\bibinfo{author}{\bibfnamefont{M.-S.} \bibnamefont{Heo}},
  \bibinfo{author}{\bibfnamefont{Y.}~\bibnamefont{Kim}},
  \bibinfo{author}{\bibfnamefont{K.}~\bibnamefont{Kim}},
  \bibinfo{author}{\bibfnamefont{G.}~\bibnamefont{Moon}},
  \bibinfo{author}{\bibfnamefont{J.}~\bibnamefont{Lee}},
  \bibinfo{author}{\bibfnamefont{H.-R.} \bibnamefont{Noh}},
  \bibinfo{author}{\bibfnamefont{M.~I.} \bibnamefont{Dykman}},
  \bibnamefont{and} \bibinfo{author}{\bibfnamefont{W.}~\bibnamefont{Jhe}},
  \bibinfo{journal}{Phys. Rev. E} \textbf{\bibinfo{volume}{82}},
  \bibinfo{pages}{031134} (\bibinfo{year}{2010}),
  \urlprefix\url{https://link.aps.org/doi/10.1103/PhysRevE.82.031134}.

\bibitem[{\citenamefont{Flicker}(2018)}]{Flicker2017}
\bibinfo{author}{\bibfnamefont{F.}~\bibnamefont{Flicker}},
  \bibinfo{journal}{SciPost Phys.} \textbf{\bibinfo{volume}{5}},
  \bibinfo{pages}{1} (\bibinfo{year}{2018}),
  \urlprefix\url{https://scipost.org/10.21468/SciPostPhys.5.1.001}.

\bibitem[{\citenamefont{Giergiel
  et~al.}(2018{\natexlab{b}})\citenamefont{Giergiel, Kuro{\'s}, and
  Sacha}}]{Giergiel2018discrete}
\bibinfo{author}{\bibfnamefont{K.}~\bibnamefont{Giergiel}},
  \bibinfo{author}{\bibfnamefont{A.}~\bibnamefont{Kuro{\'s}}},
  \bibnamefont{and} \bibinfo{author}{\bibfnamefont{K.}~\bibnamefont{Sacha}},
  \bibinfo{journal}{arXiv preprint arXiv:1807.02105}
  (\bibinfo{year}{2018}{\natexlab{b}}).

\bibitem[{\citenamefont{Guo et~al.}(2013)\citenamefont{Guo, Marthaler, and
  Sch\"on}}]{Guo2013}
\bibinfo{author}{\bibfnamefont{L.}~\bibnamefont{Guo}},
  \bibinfo{author}{\bibfnamefont{M.}~\bibnamefont{Marthaler}},
  \bibnamefont{and} \bibinfo{author}{\bibfnamefont{G.}~\bibnamefont{Sch\"on}},
  \bibinfo{journal}{Phys. Rev. Lett.} \textbf{\bibinfo{volume}{111}},
  \bibinfo{pages}{205303} (\bibinfo{year}{2013}),
  \urlprefix\url{https://link.aps.org/doi/10.1103/PhysRevLett.111.205303}.

\bibitem[{\citenamefont{Guo and Marthaler}(2016)}]{Guo2016}
\bibinfo{author}{\bibfnamefont{L.}~\bibnamefont{Guo}} \bibnamefont{and}
  \bibinfo{author}{\bibfnamefont{M.}~\bibnamefont{Marthaler}},
  \bibinfo{journal}{New Journal of Physics} \textbf{\bibinfo{volume}{18}},
  \bibinfo{pages}{023006} (\bibinfo{year}{2016}),
  \urlprefix\url{http://stacks.iop.org/1367-2630/18/i=2/a=023006}.

\bibitem[{\citenamefont{Guo et~al.}(2016)\citenamefont{Guo, Liu, and
  Marthaler}}]{Guo2016a}
\bibinfo{author}{\bibfnamefont{L.}~\bibnamefont{Guo}},
  \bibinfo{author}{\bibfnamefont{M.}~\bibnamefont{Liu}}, \bibnamefont{and}
  \bibinfo{author}{\bibfnamefont{M.}~\bibnamefont{Marthaler}},
  \bibinfo{journal}{Phys. Rev. A} \textbf{\bibinfo{volume}{93}},
  \bibinfo{pages}{053616} (\bibinfo{year}{2016}),
  \urlprefix\url{https://link.aps.org/doi/10.1103/PhysRevA.93.053616}.

\bibitem[{\citenamefont{Pengfei et~al.}(2018)\citenamefont{Pengfei, Michael,
  and Guo}}]{Liang2017}
\bibinfo{author}{\bibfnamefont{L.}~\bibnamefont{Pengfei}},
  \bibinfo{author}{\bibfnamefont{M.}~\bibnamefont{Michael}}, \bibnamefont{and}
  \bibinfo{author}{\bibfnamefont{L.}~\bibnamefont{Guo}}, \bibinfo{journal}{New
  Journal of Physics} \textbf{\bibinfo{volume}{20}}, \bibinfo{pages}{023043}
  (\bibinfo{year}{2018}), ISSN \bibinfo{issn}{1367-2630},
  \urlprefix\url{http://stacks.iop.org/1367-2630/20/i=2/a=023043}.

\bibitem[{\citenamefont{Bomantara and Gong}(2018)}]{Bomantara2018}
\bibinfo{author}{\bibfnamefont{R.~W.} \bibnamefont{Bomantara}}
  \bibnamefont{and} \bibinfo{author}{\bibfnamefont{J.}~\bibnamefont{Gong}},
  \bibinfo{journal}{Phys. Rev. Lett.} \textbf{\bibinfo{volume}{120}},
  \bibinfo{pages}{230405} (\bibinfo{year}{2018}),
  \urlprefix\url{https://link.aps.org/doi/10.1103/PhysRevLett.120.230405}.

\bibitem[{\citenamefont{Giergiel
  et~al.}(2018{\natexlab{c}})\citenamefont{Giergiel, Dauphin, Lewenstein,
  Zakrzewski, and Sacha}}]{Giergiel2018topological}
\bibinfo{author}{\bibfnamefont{K.}~\bibnamefont{Giergiel}},
  \bibinfo{author}{\bibfnamefont{A.}~\bibnamefont{Dauphin}},
  \bibinfo{author}{\bibfnamefont{M.}~\bibnamefont{Lewenstein}},
  \bibinfo{author}{\bibfnamefont{J.}~\bibnamefont{Zakrzewski}},
  \bibnamefont{and} \bibinfo{author}{\bibfnamefont{K.}~\bibnamefont{Sacha}},
  \bibinfo{journal}{arXiv preprint arXiv:1806.10536}
  (\bibinfo{year}{2018}{\natexlab{c}}).

\bibitem[{\citenamefont{Sacha}(2015{\natexlab{b}})}]{Sacha15a}
\bibinfo{author}{\bibfnamefont{K.}~\bibnamefont{Sacha}}, \bibinfo{journal}{Sci.
  Rep.} \textbf{\bibinfo{volume}{5}}, \bibinfo{pages}{10787}
  (\bibinfo{year}{2015}{\natexlab{b}}),
  \urlprefix\url{https://www.nature.com/articles/srep10787}.

\bibitem[{\citenamefont{Sacha and Delande}(2016)}]{sacha16}
\bibinfo{author}{\bibfnamefont{K.}~\bibnamefont{Sacha}} \bibnamefont{and}
  \bibinfo{author}{\bibfnamefont{D.}~\bibnamefont{Delande}},
  \bibinfo{journal}{Phys. Rev. A} \textbf{\bibinfo{volume}{94}},
  \bibinfo{pages}{023633} (\bibinfo{year}{2016}),
  \urlprefix\url{http://link.aps.org/doi/10.1103/PhysRevA.94.023633}.

\bibitem[{\citenamefont{Giergiel and Sacha}(2017)}]{Giergiel2017}
\bibinfo{author}{\bibfnamefont{K.}~\bibnamefont{Giergiel}} \bibnamefont{and}
  \bibinfo{author}{\bibfnamefont{K.}~\bibnamefont{Sacha}},
  \bibinfo{journal}{Phys. Rev. A} \textbf{\bibinfo{volume}{95}},
  \bibinfo{pages}{063402} (\bibinfo{year}{2017}),
  \urlprefix\url{https://link.aps.org/doi/10.1103/PhysRevA.95.063402}.

\bibitem[{\citenamefont{Mierzejewski et~al.}(2017)\citenamefont{Mierzejewski,
  Giergiel, and Sacha}}]{Mierzejewski2017}
\bibinfo{author}{\bibfnamefont{M.}~\bibnamefont{Mierzejewski}},
  \bibinfo{author}{\bibfnamefont{K.}~\bibnamefont{Giergiel}}, \bibnamefont{and}
  \bibinfo{author}{\bibfnamefont{K.}~\bibnamefont{Sacha}},
  \bibinfo{journal}{Phys. Rev. B} \textbf{\bibinfo{volume}{96}},
  \bibinfo{pages}{140201} (\bibinfo{year}{2017}),
  \urlprefix\url{https://link.aps.org/doi/10.1103/PhysRevB.96.140201}.

\bibitem[{\citenamefont{Delande et~al.}(2017)\citenamefont{Delande,
  Morales-Molina, and Sacha}}]{delande17}
\bibinfo{author}{\bibfnamefont{D.}~\bibnamefont{Delande}},
  \bibinfo{author}{\bibfnamefont{L.}~\bibnamefont{Morales-Molina}},
  \bibnamefont{and} \bibinfo{author}{\bibfnamefont{K.}~\bibnamefont{Sacha}},
  \bibinfo{journal}{Phys. Rev. Lett.} \textbf{\bibinfo{volume}{119}},
  \bibinfo{pages}{230404} (\bibinfo{year}{2017}),
  \urlprefix\url{https://link.aps.org/doi/10.1103/PhysRevLett.119.230404}.

\bibitem[{\citenamefont{Giergiel
  et~al.}(2018{\natexlab{d}})\citenamefont{Giergiel, Miroszewski, and
  Sacha}}]{Giergiel2017a}
\bibinfo{author}{\bibfnamefont{K.}~\bibnamefont{Giergiel}},
  \bibinfo{author}{\bibfnamefont{A.}~\bibnamefont{Miroszewski}},
  \bibnamefont{and} \bibinfo{author}{\bibfnamefont{K.}~\bibnamefont{Sacha}},
  \bibinfo{journal}{Phys. Rev. Lett.} \textbf{\bibinfo{volume}{120}},
  \bibinfo{pages}{140401} (\bibinfo{year}{2018}{\natexlab{d}}),
  \urlprefix\url{https://link.aps.org/doi/10.1103/PhysRevLett.120.140401}.

\bibitem[{\citenamefont{{Mizuta} et~al.}(2018)\citenamefont{{Mizuta},
  {Takasan}, {Nakagawa}, and {Kawakami}}}]{Mizuta2018}
\bibinfo{author}{\bibfnamefont{K.}~\bibnamefont{{Mizuta}}},
  \bibinfo{author}{\bibfnamefont{K.}~\bibnamefont{{Takasan}}},
  \bibinfo{author}{\bibfnamefont{M.}~\bibnamefont{{Nakagawa}}},
  \bibnamefont{and}
  \bibinfo{author}{\bibfnamefont{N.}~\bibnamefont{{Kawakami}}},
  \bibinfo{journal}{ArXiv e-prints}  (\bibinfo{year}{2018}),
  \eprint{1804.01291}.

\bibitem[{\citenamefont{Kosior and Sacha}(2018)}]{Kosior2017}
\bibinfo{author}{\bibfnamefont{A.}~\bibnamefont{Kosior}} \bibnamefont{and}
  \bibinfo{author}{\bibfnamefont{K.}~\bibnamefont{Sacha}},
  \bibinfo{journal}{Phys. Rev. A} \textbf{\bibinfo{volume}{97}},
  \bibinfo{pages}{053621} (\bibinfo{year}{2018}),
  \urlprefix\url{https://link.aps.org/doi/10.1103/PhysRevA.97.053621}.

\bibitem[{\citenamefont{D\'ora et~al.}(2013)\citenamefont{D\'ora, Pollmann,
  Fort\'agh, and Zar\'and}}]{Dora2013}
\bibinfo{author}{\bibfnamefont{B.}~\bibnamefont{D\'ora}},
  \bibinfo{author}{\bibfnamefont{F.}~\bibnamefont{Pollmann}},
  \bibinfo{author}{\bibfnamefont{J.}~\bibnamefont{Fort\'agh}},
  \bibnamefont{and} \bibinfo{author}{\bibfnamefont{G.}~\bibnamefont{Zar\'and}},
  \bibinfo{journal}{Phys. Rev. Lett.} \textbf{\bibinfo{volume}{111}},
  \bibinfo{pages}{046402} (\bibinfo{year}{2013}),
  \urlprefix\url{https://link.aps.org/doi/10.1103/PhysRevLett.111.046402}.

\bibitem[{\citenamefont{Canovi et~al.}(2014)\citenamefont{Canovi, Werner, and
  Eckstein}}]{Canovi2014}
\bibinfo{author}{\bibfnamefont{E.}~\bibnamefont{Canovi}},
  \bibinfo{author}{\bibfnamefont{P.}~\bibnamefont{Werner}}, \bibnamefont{and}
  \bibinfo{author}{\bibfnamefont{M.}~\bibnamefont{Eckstein}},
  \bibinfo{journal}{Phys. Rev. Lett.} \textbf{\bibinfo{volume}{113}},
  \bibinfo{pages}{265702} (\bibinfo{year}{2014}),
  \urlprefix\url{https://link.aps.org/doi/10.1103/PhysRevLett.113.265702}.

\bibitem[{\citenamefont{Fogarty et~al.}(2017)\citenamefont{Fogarty, Usui,
  Busch, Silva, and Goold}}]{Fogarty2017}
\bibinfo{author}{\bibfnamefont{T.}~\bibnamefont{Fogarty}},
  \bibinfo{author}{\bibfnamefont{A.}~\bibnamefont{Usui}},
  \bibinfo{author}{\bibfnamefont{T.}~\bibnamefont{Busch}},
  \bibinfo{author}{\bibfnamefont{A.}~\bibnamefont{Silva}}, \bibnamefont{and}
  \bibinfo{author}{\bibfnamefont{J.}~\bibnamefont{Goold}},
  \bibinfo{journal}{New Journal of Physics} \textbf{\bibinfo{volume}{19}},
  \bibinfo{pages}{113018} (\bibinfo{year}{2017}),
  \urlprefix\url{http://stacks.iop.org/1367-2630/19/i=11/a=113018}.

\bibitem[{\citenamefont{Zi\ifmmode~\acute{n}\else \'{n}\fi{}
  et~al.}(2008)\citenamefont{Zi\ifmmode~\acute{n}\else \'{n}\fi{},
  Ole\ifmmode~\acute{s}\else \'{s}\fi{}, Trippenbach, and Sacha}}]{Zin2008a}
\bibinfo{author}{\bibfnamefont{P.}~\bibnamefont{Zi\ifmmode~\acute{n}\else
  \'{n}\fi{}}},
  \bibinfo{author}{\bibfnamefont{B.}~\bibnamefont{Ole\ifmmode~\acute{s}\else
  \'{s}\fi{}}}, \bibinfo{author}{\bibfnamefont{M.}~\bibnamefont{Trippenbach}},
  \bibnamefont{and} \bibinfo{author}{\bibfnamefont{K.}~\bibnamefont{Sacha}},
  \bibinfo{journal}{Phys. Rev. A} \textbf{\bibinfo{volume}{78}},
  \bibinfo{pages}{023620} (\bibinfo{year}{2008}),
  \urlprefix\url{https://link.aps.org/doi/10.1103/PhysRevA.78.023620}.

\bibitem[{\citenamefont{Pethick and Smith}(2002)}]{Pethick2002}
\bibinfo{author}{\bibfnamefont{C.}~\bibnamefont{Pethick}} \bibnamefont{and}
  \bibinfo{author}{\bibfnamefont{H.}~\bibnamefont{Smith}},
  \emph{\bibinfo{title}{{Bose-Eistein condensation in dilute gases}}}
  (\bibinfo{publisher}{{Cambridge University Press}},
  \bibinfo{address}{{Cambridge, England}}, \bibinfo{year}{2002}).

\bibitem[{\citenamefont{Carr et~al.}(2000)\citenamefont{Carr, Clark, and
  Reinhardt}}]{Carr2000}
\bibinfo{author}{\bibfnamefont{L.~D.} \bibnamefont{Carr}},
  \bibinfo{author}{\bibfnamefont{C.~W.} \bibnamefont{Clark}}, \bibnamefont{and}
  \bibinfo{author}{\bibfnamefont{W.~P.} \bibnamefont{Reinhardt}},
  \bibinfo{journal}{Phys. Rev. A} \textbf{\bibinfo{volume}{62}},
  \bibinfo{pages}{063611} (\bibinfo{year}{2000}),
  \urlprefix\url{http://link.aps.org/doi/10.1103/PhysRevA.62.063611}.

\bibitem[{\citenamefont{Lebedev and Silverman}(1965)}]{Lebedev1965}
\bibinfo{author}{\bibfnamefont{N.~N.} \bibnamefont{Lebedev}} \bibnamefont{and}
  \bibinfo{author}{\bibfnamefont{R.~A.} \bibnamefont{Silverman}},
  \emph{\bibinfo{title}{{Special functions and their applications}}}
  (\bibinfo{publisher}{Prentice-Hall}, \bibinfo{address}{Englewood Cliffs,
  N.J.}, \bibinfo{year}{1965}).

\bibitem[{Ryz(2015)}]{Ryzhik2015}
in \emph{\bibinfo{booktitle}{Table of Integrals, Series, and Products (Eighth
  Edition)}}, edited by
  \bibinfo{editor}{\bibfnamefont{D.}~\bibnamefont{Zwillinger}},
  \bibinfo{editor}{\bibfnamefont{V.}~\bibnamefont{Moll}},
  \bibinfo{editor}{\bibfnamefont{I.}~\bibnamefont{Gradshteyn}}, ,
  \bibnamefont{and} \bibinfo{editor}{\bibfnamefont{I.}~\bibnamefont{Ryzhik}}
  (\bibinfo{publisher}{Academic Press}, \bibinfo{address}{Boston},
  \bibinfo{year}{2015}), pp. \bibinfo{pages}{249 -- 519},
  \bibinfo{edition}{eighth edition} ed., ISBN
  \bibinfo{isbn}{978-0-12-384933-5},
  \urlprefix\url{https://www.sciencedirect.com/science/article/pii/B9780123849335000035}.

\end{thebibliography}

\end{document}